\documentclass[sn-basic]{sn-jnl}

\usepackage{graphicx}%
\usepackage{multirow}%
\usepackage{amsmath,amssymb,amsfonts}%
\usepackage{amsthm}%
\usepackage{mathrsfs}%
\usepackage[title]{appendix}%
\usepackage{xcolor}%
\usepackage{textcomp}%
\usepackage{manyfoot}%
\usepackage{booktabs}%
\usepackage{algorithm}%
\usepackage{algorithmicx}%
\usepackage{algpseudocode}%
\usepackage{listings}%
\usepackage{dirtytalk}

\usepackage[utf8]{inputenc}
\usepackage{cite}
\usepackage{graphicx}
\usepackage{caption}
\usepackage{subcaption}

\theoremstyle{thmstyleone}%
\theoremstyle{thmstyletwo}%
\theoremstyle{thmstylethree}%

\begin{document}

\title[Fokas-Lenells Soliton Management and Landau-Lifshitz Equation]{Soliton Management for ultrashort pulse: dark  and anti-dark solitons of Fokas-Lenells equation with a damping like perturbation and a gauge equivalent spin system}

\author[1]{\fnm{Riki} \sur{Dutta}}\email{rikidutta96@gmail.com}

\author[1]{\fnm{Gautam K.} \sur{Saharia}}\email{gautamks0802@gmail.com}
\equalcont{These authors contributed equally to this work.}

\author[1]{\fnm{Sagardeep} \sur{Talukdar}}\email{talukdarsagardeep@gmail.com}
\equalcont{These authors contributed equally to this work.}

\author*[1]{\fnm{Sudipta} \sur{Nandy}}\email{sudipta.nandy@cottonuniversity.ac.in}
\equalcont{These authors contributed equally to this work.}

\affil[1]{\orgdiv{Department of Physics}, \orgname{Cotton University}, \orgaddress{\street{Panbazar}, \city{Guwahati}, \postcode{781001}, \state{Assam}, \country{India}}}

\abstract{We investigate the propagation of an ultrashort optical pulse using Fokas-Lenells equation (FLE) under varying dispersion, nonlinear effects and perturbation. Such a system can be said to be under soliton management (SM) scheme. At first, under a gauge transformation, followed by shifting of variables, we transform FLE under SM into a simplified form, which is similar to an equation given by Davydova and Lashkin for plasma waves, we refer to this form as DLFLE. Then, we propose a  bilinearization for DLFLE in a non-vanishing background by
introducing an auxiliary function which transforms DLFLE into three bilinear equations. We solve these equations and obtain dark and anti-dark one-soliton solution (1SS) of DLFLE. From here, by reverse transformation of the solution, we obtain the 1SS of FLE and explore the soliton behavior under different SM schemes. Thereafter, we obtain dark and anti-dark two-soliton solution (2SS) of DLFLE and determine the shift in phase of the individual solitons on interaction through asymptotic analysis. We then, obtain the 2SS of FLE and represent the soliton graph for different SM scheme. Thereafter, we present the procedure to determine N-soliton solution (NSS) of DLFLE and FLE. Later, we introduce a Lax pair for DLFLE and through a gauge transformation we convert the spectral problem of our system into that of an equivalent spin system which is termed as Landau-Lifshitz (LL) system. LL equation (LLE) holds the potential to provide information about various nonlinear structures and properties of the system.}


\maketitle
\section{Introduction}\label{sec1}

In optics, soliton is a nonlinear optical pulse that propagates through a waveguide without getting distorted and hence can acts as a good information carrier in digital and quantum communication devices. So the research on soliton has always been in great demand. For the soliton to exist the system must balance the dispersion and nonlinear effects. Nonlinear Schrödinger equation (NLSE) is the most basic equation that addresses both of the two effects and solving it results in a soliton solution \citep{hasegawa1973transmission, hasegawa1973transmission2, malomed1996soliton, hosseini2023dynamics, serkin2000novel, chakraborty2015bilinearization}. Beyond optical soliton, NLSE (popularly known by the name Gross–Pitaevskii equation \citep{pethick2008bose}) also describes the propagation of matter-wave in Bose-Einstein condensates, where the dispersion is represented by kinetic energy of the bosons and nonlinear effect arises due to intra and inter bosonic interactions. In optics, besides NLSE, there are other equations taking into account some higher order effects. Some of these well-studied equations are Kaup--Newell equation (KNE) \citep{jawad2019bright}, Chen–Lee–Liu equation (CLLE) \citep{gonzalez2018w, yildirim2020optical}, Tiriki-Biswas equation (TBE) \citep{triki2018sub}, sasa-satsuma equation (SSE) \citep{liu2017certain, hosseini2022generalized, yildirim2019opticalss}, etc. It may be interesting to point out that KNE, CLLE and TBE are also referred to as byproducts of derivative NLSE (DNLSE). However, a high intensity light source can produce ultrashort (femtosecond) pulses. To describe such a pulse, one must account for the higher order effects namely, spatio-temporal dispersion (STD) and nonlinear dispersion (ND) effect along with the group velocity dispersion (GVD) and Kerr nonlinear effects (KNLE). Fokas-Lenells equation (FLE) \citep{lenells2008novel, lenells2009exactly} is a nonlinear equation that addresses all these effects. The dimensionless form of FLE is presented as
\begin{align} 
	\label{VFLE}
	iU_t  + a_1 U_{xx} - a_2 U_{xt} + b |U|^2\ ( U + i\ a_2 U_x) &= 0 
\end{align}
where $U$ is the field function that describes the complex waveform of an ultrashort pulse. The suffix $x$ and $t$ denote the partial differentiations of $U$ with respect to $x$ and $t$ respectively. $U_t$ is the temporal evolution of the pulse, $U_{xx}$, $U_{xt}$, $|U|^2 U$ and $|U|^2 U_x$ represents GVD, STD, KNLE and ND respectively. Eqn. (\ref{VFLE}) is relatively a new equation and the study on this is going on in full swing. Many analytical approaches have been employed to derive soliton solutions of FLE \citep{biswas2018optical, hosseini2020optical, triki2017combined, baronio2015baseband, chen2018peregrine, cinar2022derivation, onder2022obtaining, ullah2023optical, gomez2022soliton, el2023novel, gaballah2023generalized, krishnan2019optical, matsuno2012direct, matsuno2012direct1, talukdar2023multi}. In general the works on FLE is done for constant dispersion and nonlinear coefficients. Only a few papers have been published with variable coefficients in FLE \citep{kundu2010two, lu2013nonautonomous, wang2017higher}. In this manuscript, however, our purpose is to study a system of an ultrashort pulse having time-varying dispersion, nonlinear effects and under a damping like perturbation. Under the SM scheme, Eqn. (\ref{VFLE}) modified accordingly and rewritten as
\begin{align}
	\label{FLE}
	iU_t  + a_1 D(t)\ U_{xx} - a_2 U_{xt} + b R(t) |U|^2\ (U + i\ a_2 U_x) &= \Gamma(t)\ (U_x - i\ n\ U)
\end{align}
here the time-variable coefficient parameters $D(t)$ and $R(t)$ represents variation in dispersion and nonlinear effects respectively and $\Gamma(t)$ is the gain parameter. The form of $\Gamma(t)$ in this manuscript is considered as $\Gamma(t) = \frac{R(t) D^{\prime}(t) - R^{\prime}(t) D(t)}{2 D(t) R(t)}$ and the symbol prime ($^{\prime}$) represents the derivative with respect to $t$. The first and second terms on the right hand side of Eqn. (\ref{FLE}) are for damping and gain (or loss) of the medium. From hereon throughout the manuscript wherever we mention FLE, we are referring to Eqn. (\ref{FLE}). The two main objectives of our manuscript are 1.) at first, we introduce a suitable bilinear scheme to obtain dark and anti-dark soliton solutions of Eqn. (\ref{FLE}). Thereafter, explore the soliton solution under different SM schemes and provide a systematic procedure to evaluate higher order soliton solutions. 2.) later, we are going to transform an already gauge transformed form of Eqn. (\ref{FLE}) which we call as DLFLE (Eqn. (\ref{DLFLE})), into an equation of equivalent spin system called LLE. To the best of our knowledge this work had not been published before for such a system. \\

To address the first objective, we at first, through a gauge transformation followed by change in variables, convert Eqn. (\ref{FLE}) into a simplified form (DLFLE), then we propose a bilinear scheme and introduce an auxiliary function to make the bilinear process more convenient. Our proposed scheme transforms the higher order nonlinear equation into three bilinear forms. We obtain dark and anti-dark 1SS of DLFLE. Now, by reverse transforming the obtained 1SS, we get the 1SS of FLE. Then we mention the criteria upon which the nature of soliton becomes dark or anti-dark and also provide the graphical representation under different SM schemes. We also obtain 2SS of DLFLE and determine the shift in phase of the two individual solitons upon interaction. Eventually, we obtain the 2SS of FLE. In our previous paper \citep{dutta2023fokas}, we had implemented an analogous scheme to realize soliton solution of FLE (DLFLE form to be precise), but for constant coefficients ($D(t)=R(t)=1$) and now we extend the scheme to realize the soliton of FLE under SM (Eqn. (\ref{FLE})). \\

Again, it is observed that various nonlinear equations shows gauge equivalence with spin system equations (referred as LLE) \citep{takhtajan1979equivalence, kundu1984landau, ghosh1999soliton, ghosh1999inverse}, even DLFLE having $D(t) = R(t) =1$ shows gauge equivalence LLE \citep{dutta2023fokas}. LLE can provide interesting features about a system which are important from  physical and mathematical point of view \citep{lakshmanan2011fascinating, guo2008landau}. We now introduce a Lax pair for DLFLE and from that to accomplish our second objective, we are going to search for a gauge transformation which transforms the spectral problem of DLFLE into a spectral problem of LLE. \\

This manuscript is arranged as: in the following section we consider a gauge transformation followed by change in variables that will convert Eqn. (\ref{FLE}) into DLFLE (Eqn. (\ref{DLFLE})). Then using Hirota bilinearization, we derive dark and anti-dark 1SS of DLFLE and eventually 1SS of FLE and present the graphs under various SM scheme. Then, derive the multi-soliton solutions and perform asymptotic analysis. In the third section we propose a Lax pair for DLFLE and through a gauge transformation we transform the system into a gauge equivalent LLE. The fourth section will conclude the manuscript.

\section{Bilinearization of FLE with non-vanishing background}\label{sec2}

The transformation of Eqn. (\ref{FLE}) with respect to the variables ($x$ and $t$) gives us the liberty to choose the constant coefficients ($a_1$, $a_2$, $b$ and $n$) as per our convenience since the coefficients can be absorbed in the process of variable transformation. Assuming $n=\frac{1}{a_2} $, $m=\frac{a_1}{a_2} > 0$ and positive $b$, consider the gauge transformation 
\begin{align}
	\label{GT}
	U = \sqrt{\frac{m}{b}}n e^{i(n x +2 mn t)} u 
\end{align}
followed by the transformation of variables
\begin{align}
	\label{VT}
	\xi = 2(x+m t), \quad \tau=-\frac{mn^2}{2} t
\end{align}
and for the time-varying parameters
\begin{align}
	D(t) \rightarrow D(\tau), \quad R(t) \rightarrow R(\tau), \quad \Gamma(t) \rightarrow \Gamma(\tau) \nonumber
\end{align}
we get the following equation
\begin{align} 
	\label{DLFLE}
	u_{\xi \tau} - D(\tau)\ u + 2i\ R(\tau)\ |u|^2 u_\xi &= \Gamma(\tau)\ u_x
\end{align}
here $u$ is the field function corresponding to the new transformed system. For constant dispersion and nonlinear effects, i.e. constant coefficients ($D(t)=R(t)=1$), Eqn. (\ref{DLFLE}) is the first negative hierarchy of DNLSE and is similar to an equation governing the dynamics of short-wavelength ion-cyclotron waves in plasma \citep{davydova1991short, lashkin2021perturbation}. Davydova and Lashkin first realized this and hence in case of constant coefficients, Eqn. (\ref{DLFLE}) can be termed as Davydova Lashkin FLE (DLFLE). In this manuscript, we refer Eqn. (\ref{DLFLE}) as DLFLE, even though $D(t)$ and $R(t)$ are not constants. For a constant $\Gamma(\tau) = \Gamma$, the right hand side of Eqn. (\ref{DLFLE}) takes the form of a perturbation representing linear damping \citep{berger1976thresholds}. \\

For a soliton solution of DLFLE, we assume a non-vanishing background condition $u \rightarrow \rho\ e^{i \ (\kappa \xi + \omega(\tau))}$ as $ \xi \rightarrow \pm \infty $. Under this situation, we expect dark and anti-dark soliton solutions of Eqn. (\ref{DLFLE}) through bilinearization.

\vspace{2mm}
To write  Eqn. (\ref{DLFLE}) in the bilinear form let us assume
\begin{align}
	\label{bilin0}
	u& = \sqrt{\frac{D(\tau)}{R(\tau)}}\ \frac{g}{f}
\end{align}
where $g$ and  $f$ are two complex functions of ($\xi, \tau$). Consequently, Eqn. (\ref{DLFLE}) becomes
\begin{multline}
	\frac{1}{f^2}(D_\xi D_\tau - D(\tau)) g.f - \frac{g}{f^3}D_\xi D_\tau(f.f) +  \frac{2i\ |g|^2}{f^3 f^*} D(\tau) D_\xi (g.f) + \frac{g \lambda f.f}{f^3} - \frac{\lambda g.f}{f^2} + \\
	\label{Bilin}
	\frac{s |g|^2}{f^3} - \frac{s |g|^2 f^*}{f^3 f^*} =0
\end{multline} 
where $D_\xi$, $D_\tau $ are  Hirota derivatives \citep{hirota1976n} and are defined as
\begin{align}
	D_\xi^m D_\tau^n g(\xi,\tau).f(\xi,\tau)= 
	(\frac{\partial}{\partial \xi}  - \frac{\partial}{\partial \xi^\prime})^m
	(\frac{\partial}{\partial \tau}  - \frac{\partial}{\partial \tau^\prime})^n
	g(\xi,\tau).f(\xi^\prime,\tau^\prime)\Bigg|_{ (\xi=\xi^\prime)(\tau= \tau^\prime)}
\end{align} 

Notice that the last two terms in Eqn. (\ref{Bilin}) contain an  auxiliary function $s$  which is introduced so that the multilinear Eqn. (\ref{Bilin}) can be cast into three bilinear equations given below (Eqns. (\ref{BR1}) - (\ref{BR3})). Here, the constant $\lambda$ will also be determined while solving the three equations. The bilinear equations in terms of $g$, $f$ and $s$  are 
\begin{align}
	\label{BR1}
	(D_{\xi} D_{\tau} - D(\tau) - \lambda)g.f&=0 \\
	\label{BR2}
	(D_{\xi} D_{\tau}  - \lambda)f.f  &= sg^* \\
	\label{BR3}
	2i\ D(\tau) D_{\xi} (g.f)   &= sf^* 
\end{align}

Now to obtain soliton solution, the expansions of $g$, $f$ and $s$  with respect to an arbitrary parameter $\epsilon$ are as follows
\begin{align}
	\label{GF}
	g&= g_0 \ (1 + \epsilon^2 g_2 + \epsilon^4 g_4 + ... ), \quad \quad \quad 
	f= 1 + \epsilon^2 f_2 + \epsilon^4 f_4 + ...\\
	\label{S}
	s& = s_0 \ (1 + \epsilon^2 s_2 + \epsilon^4 s_4 + ... )
\end{align}  

\subsection{Dark and Anti-Dark 1SS}\label{subsec2.1}

To obtain the 1SS of DLFLE, we drop the terms of order greater than or equal to  $\epsilon^3 $ in  $g$, $f$ and $s$. Thus, Eqn. (\ref{bilin0}) can be represented as
\begin{align}
	\label{sol1}
	u &= \sqrt{\frac{D(\tau)}{R(\tau)}}\ \frac{g_0 (1 + \epsilon^2 g_2^{(1)})}{1 + \epsilon^2 f_2^{(1)}}\Big|_{\epsilon=1}
\end{align}
Let us consider the expressions for $g_0$, $s_0$, $g_2^{(1)}$, $s_2^{(1)}$ and $f_2^{(1)}$ are as follows
\begin{align}
	\label{g0}
	g_0 &= \rho \ e^{i \ (\kappa \xi + \omega(\tau))}\\
	\label{s0}
	s_0 &= \rho_s \ e^{i \ (\kappa \xi + \omega(\tau))} \\
	\label{g2}
	g_2^{(1)} &= K_1 \ e^{\theta_1 + \theta_1^*}\\
	\label{s2}
	s_2^{(1)} &= M_1 \ e^{\theta_1 + \theta_1^*} \\
	\label{f2}
	f_2^{(1)} &= T_1 \ e^{\theta_1 + \theta_1^*}
\end{align}
where $\theta_1$ = $p_1 \ \xi + \Omega_1 \int D(\tau) d\tau$. $p_1$, $\Omega_1$, $K_1$, $M_1$, $T_1$ are complex parameters. Let us consider, $p_1$ = $p_{1r} + i\ p_{1i}$ and $T_1$ = $T_{1r} + i\ T_{1i}$ where $p_{1r}$, $p_{1i}$, $T_{1r}$ and $T_{1i}$ are real. On substituting the above expressions into Eqns. (\ref{BR1}) - (\ref{BR3}) yield the following
\begin{align}
	\label{rhos}
	\rho_s &= - 2\ \kappa \rho\\
	\label{lambda}
	\lambda &= 2\ \kappa \rho^2\\
	\label{omega}
	\omega(\tau)  &= (-\frac{1}{\kappa} - 2\ \rho^2) \int D(\tau) d\tau  \\
	\label{Omega}
	\Omega_1 &= \frac{h_1}{p_1}\\
	\label{m}
	M_1 &= \frac{T_1^2}{K_1^*}\\
	\label{k}
	K_1 &= \gamma_1 \ T_1^*
\end{align}
where $h_1$ is real and $\gamma_1$ is a complex of absolute value 1, and these two are represented as
\begin{align}
	\label{gamma}
	\gamma_1 &= \frac{(p_1 + p_1^*) T_1 - i (T_1 - T_1^*) \kappa}{(p_1 + p_1^*) T_1^* - i (T_1 - T_1^*) \kappa}\\
	\label{h}
	h_1 &= -\frac{|p_1|^2\ (-1 + \gamma_1)^2\ \ \kappa \rho^2}{(p_1 + p_1^*)^2 \gamma_1}
\end{align}
and the system obeys the constraint
\begin{align}
	\label{cons}
	p_{1r}^2 \ |T_1|^2 + 2 p_{1r} \ \kappa \ (1 + \kappa \rho^2)\ T_{1r} \ T_{1i} + \kappa^2 \ (1 + \kappa \rho^2)\ T_{1i}^2 &= 0
\end{align}
now keeping either $T_{1r}$ or $T_{1i}$ fixed, we can calculate the other from Eqn. (\ref{cons}). And this will also lead to a restriction $\ |p_{1r}|\ \le\ \sqrt{ \kappa^3 \rho^2\ (1 + \kappa \rho^2)}$. It is interesting to point out that in case of $D(t) = R(t) = 1$, the obtained expressions for $g$, $f$ align with the results already deduced for DLFLE having constant coefficients \citep{dutta2023fokas}, which is expected. \\

As already mentioned, the soliton obtained will always have a background of $\rho$, and the nature of soliton being dark (dip in the background) or anti-dark (bulge on the background) depends upon the parameters $p_{1r}$, $\kappa$ and $T_{1i}$. For positive $p_{1r}$, when $\kappa$ and $T_{1i}$ have different signs we get a dark soliton and for the same sign of $\kappa$ and $T_{1i}$ the soliton becomes anti-dark. For the case of negative $p_{1r}$ the criteria become vice versa. This statement is supported by the expression of the amplitude ($A_1$) given as
\begin{align}
	\label{Amp1}
	A_1 &= \rho\ \Big|\sqrt{\frac{D(\tau)}{R(\tau)}}\ \frac{T_1 + \gamma_1\  |T_1|}{T_1 + |T_1|}\Big|
\end{align}

For the mentioned criteria of anti-dark soliton in the above paragraph, $A_1$ is greater than $\rho$ and for the dark soliton criteria, $A_1$ is less than $\rho$. Now, the obtained 1SS \say{$u$} (Eqn. (\ref{sol1})) is a solution of DLFLE). By replacing $\xi$ and $\tau$ in terms of $x$ and $t$ from Eqn. (\ref{VT}) and putting the expression of \say{$u$} inside of \say{$U$} in Eqn. (\ref{GT}), we get the 1SS of FLE. One important and obvious detail we like to mention at this point that all the anti-dark 1SS of FLE presented in Figs. (\ref{Fig1}) - (\ref{Fig6}) have parameters $p_{1r} = -10$, $\kappa = 5$, $T_{1r} = -6$ and for all the dark 1SS graphs, we have $p_{1r} = 10$, $\kappa = -5$, $T_{1r} = 6$. $T_{1r}$ is calculated using Eqn. (\ref{cons}). And for all the graphs, we fix $\rho = 2$, $m = 0.2$, $n = 4$, $b = 1$, $a_2 = \frac{1}{n}$, $a_1 = m\ a_2$.\\

Again the role of $D(\tau)$, $R(\tau)$ and $\Gamma(\tau)$ very much impact the structure and propagation of the solitons. The existence of $\Gamma(\tau)$ also affects the background of the system. Depending upon the relationship between $D(\tau)$ and $R(\tau)$, $\Gamma(\tau)$ either vanishes or exists. In the following subsections (\ref{subsubsec2.1.1}) - (\ref{subsubsec2.1.5}) we are going to drive into some of the different SM scheme.

\subsubsection{$D(t) = R(t) = 1$ (or some other constants) and $\Gamma(t)$ vanishes}\label{subsubsec2.1.1}

\begin{figure}
	\centering
	\begin{subfigure}{.45\textwidth}
		\includegraphics[width=\textwidth]{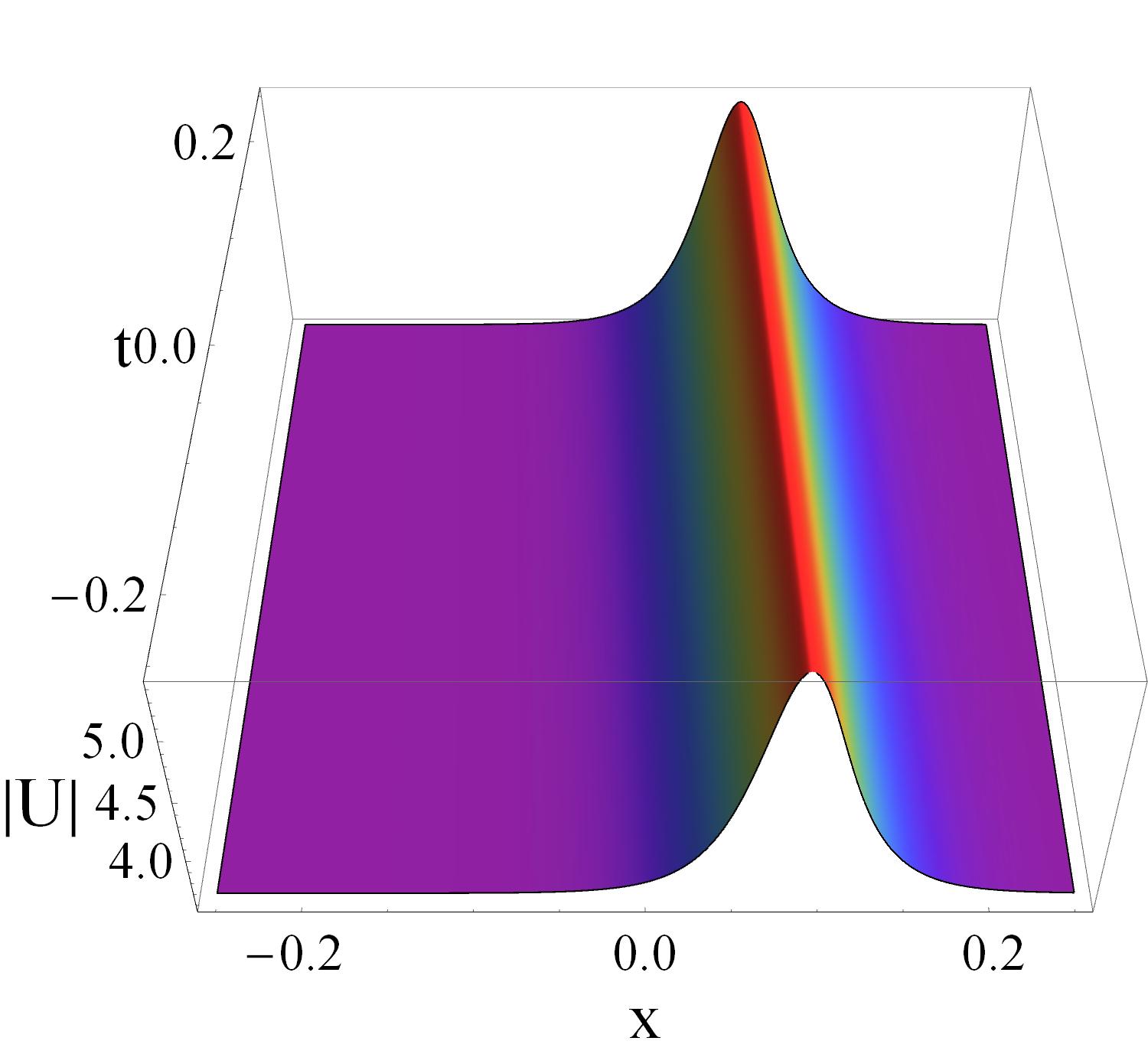}
		\caption{Anti-Dark 1SS}
		\label{Fig1a}
	\end{subfigure}
	\begin{subfigure}{.45\textwidth}
		\includegraphics[width=\textwidth]{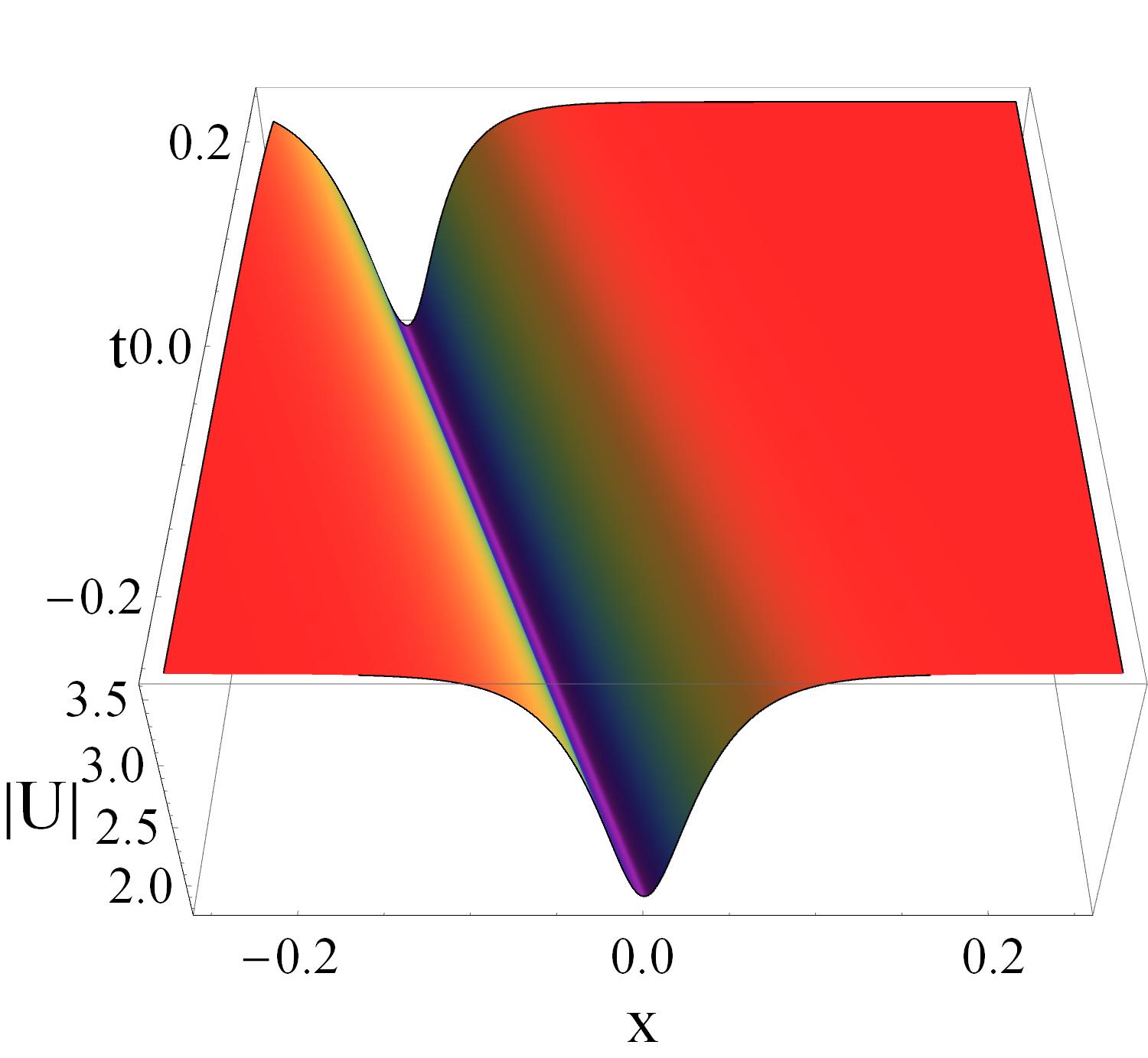}
		\caption{Dark 1SS}
		\label{Fig1b}
	\end{subfigure}
	\caption{3D plot representation of 1SS under SM scheme: $D(t) = R(t) = 1$. (a) represents anti-dark soliton and (b) represents dark soliton.}
	\label{Fig1}
\end{figure}

$D(t)$ and $R(t)$ are linearly dependent functions, hence $\Gamma(t) = 0$. The background in this case remains unaffected since the gain parameter is zero. Fig. (\ref{Fig1a}) demonstrates this condition for anti-dark 1SS in a 3D plot. Fig. (\ref{Fig1b}) demonstrates the same but for dark 1SS.

\subsubsection{$D(t) = 1$, $R(t) = 1 + \sigma cos(kt)$ and $\Gamma(t)$ is non-vanishing}\label{subsubsec2.1.2}

\begin{figure}
	\centering
	\begin{subfigure}{.3\textwidth}
		\includegraphics[width=\textwidth]{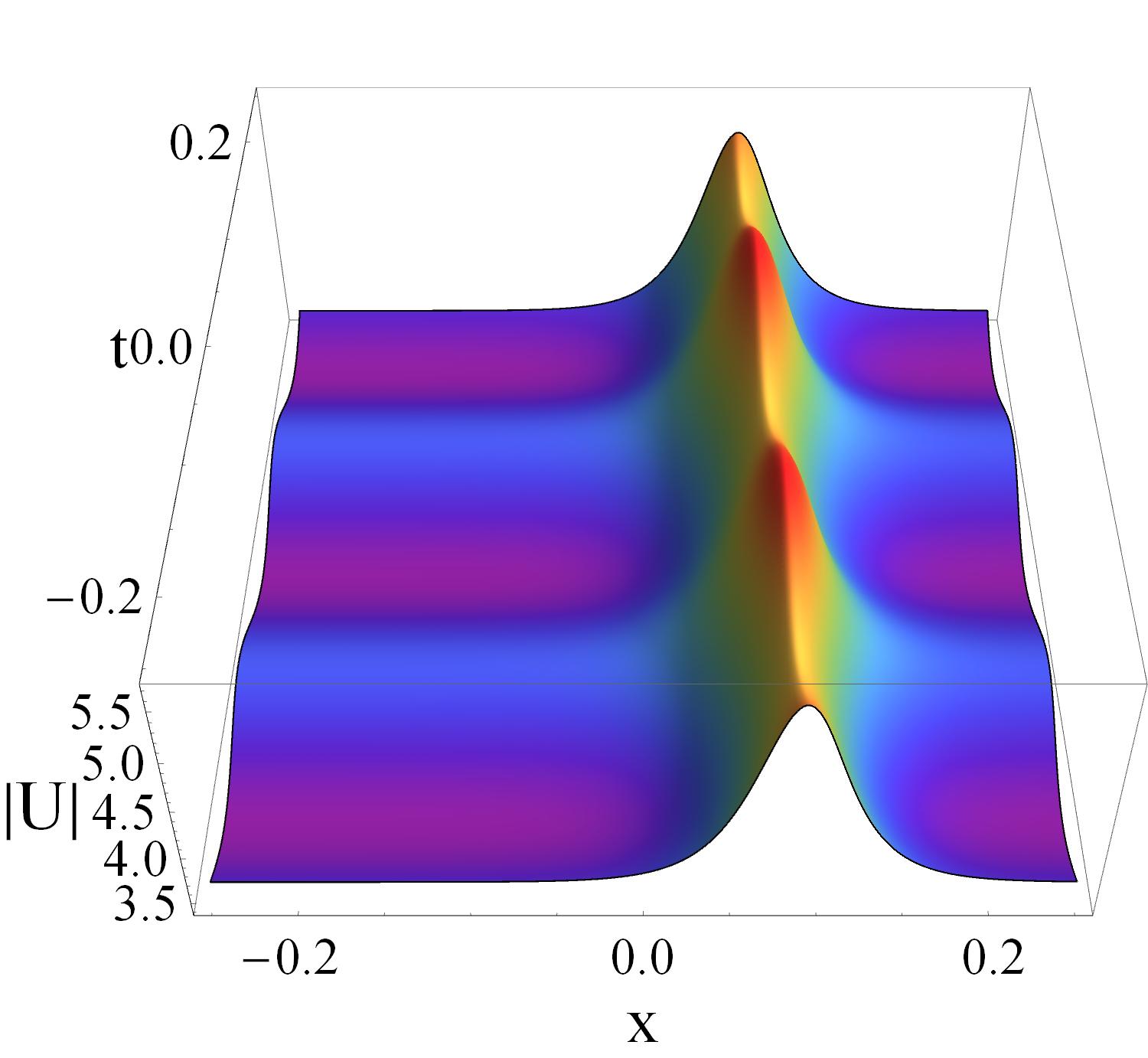}
		\caption{Anti-Dark 1SS}
		\label{Fig2a}
	\end{subfigure}
	\begin{subfigure}{.3\textwidth}
		\includegraphics[width=\textwidth]{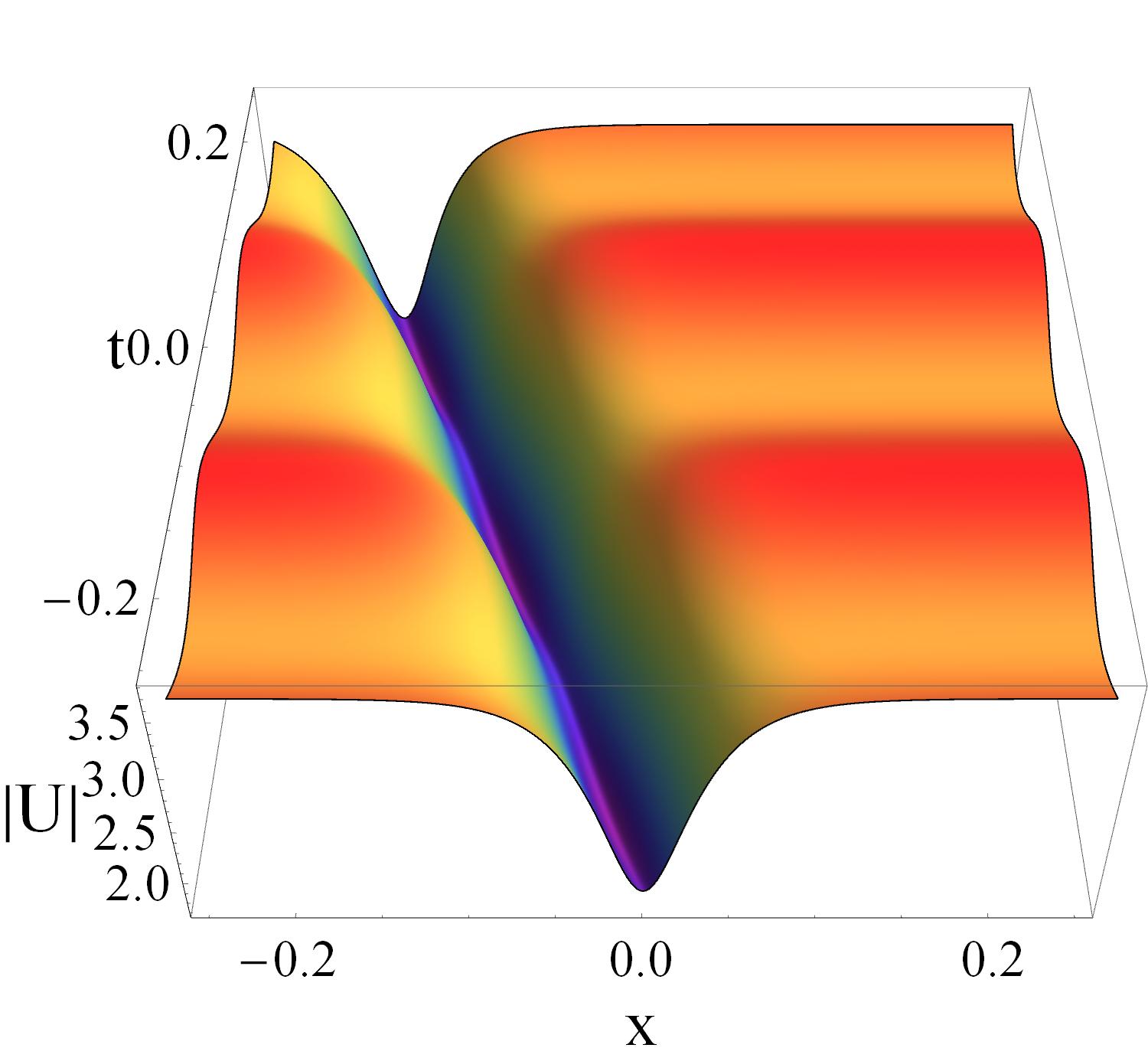}
		\caption{Dark 1SS}
		\label{Fig2b}
	\end{subfigure}
	\begin{subfigure}{.3\textwidth}
		\includegraphics[width=\textwidth]{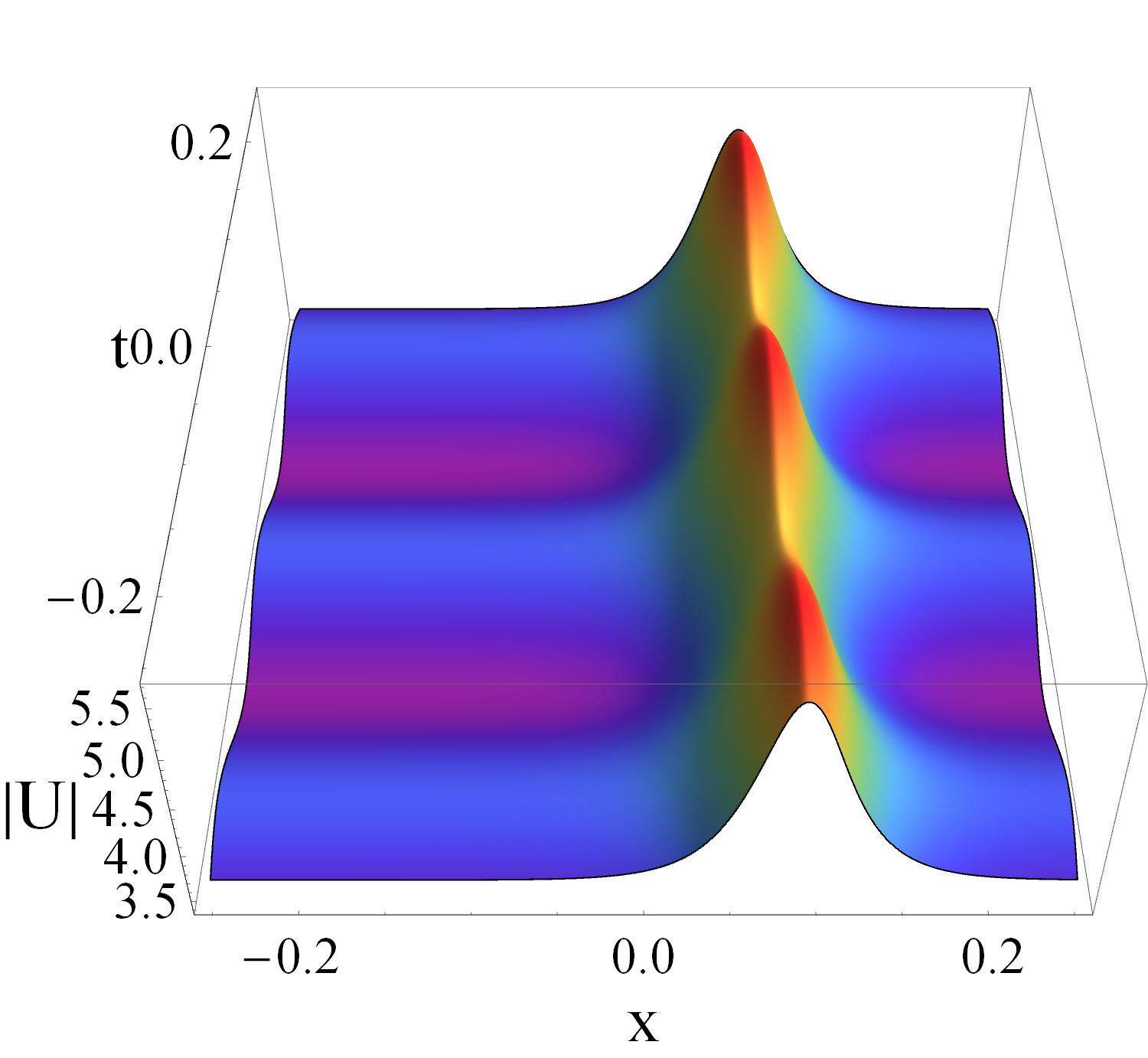}
		\caption{Anti-Dark 1SS}
		\label{Fig2c}
	\end{subfigure}
	\caption{3D plot representation of 1SS under SM scheme: $D(t) = 1$, $R(t) = 1 + \sigma cos(kt)$ with $\sigma = 0.1$ and $k = 10 \pi$. (a) represents anti-dark soliton and (b) represents dark soliton. (c) represents anti-dark soliton but under SM scheme $D(t) = 1 + \sigma cos(kt)$, $R(t) = 1$.}
	\label{Fig2}
\end{figure}

$D(t)$ and $R(t)$ are linearly independent functions, hence $\Gamma(t) \ne 0$. The background in this case becomes sinusoidal for the contribution of \say{cosine} function in $R(t)$ and the gain parameter being non-zero. Figs. (\ref{Fig2a}) - (\ref{Fig2b}) demonstrate this condition. We keep the value of $\sigma$ below 1 to avoid the scenario $R(t) \to 0$ and eventually $u, U \to \infty$, which can be problematic in graphical presentation. If $D(t)$ and $R(t)$ interchange their values than the sinusoidal nature of the graph reserves (ups and downs get interchanged) as demonstrated in Fig. (\ref{Fig2c}). And for the same expressions of $D(t)$ and $R(t)$ (= \say{$1 + \sigma cos(kt)$}), implies $\Gamma(t) = 0$, then for a small value of $\sigma$, the graph will become similar to that of Figs. (\ref{Fig1a}) and (\ref{Fig1b}).

\subsubsection{$D(t) = 1$, $R(t) = 1 + \sigma e^{(-k t^2)}$ and $\Gamma(t)$ is non-vanishing}\label{subsubsec2.1.3}

\begin{figure}
	\centering
	\begin{subfigure}{.3\textwidth}
		\includegraphics[width=\textwidth]{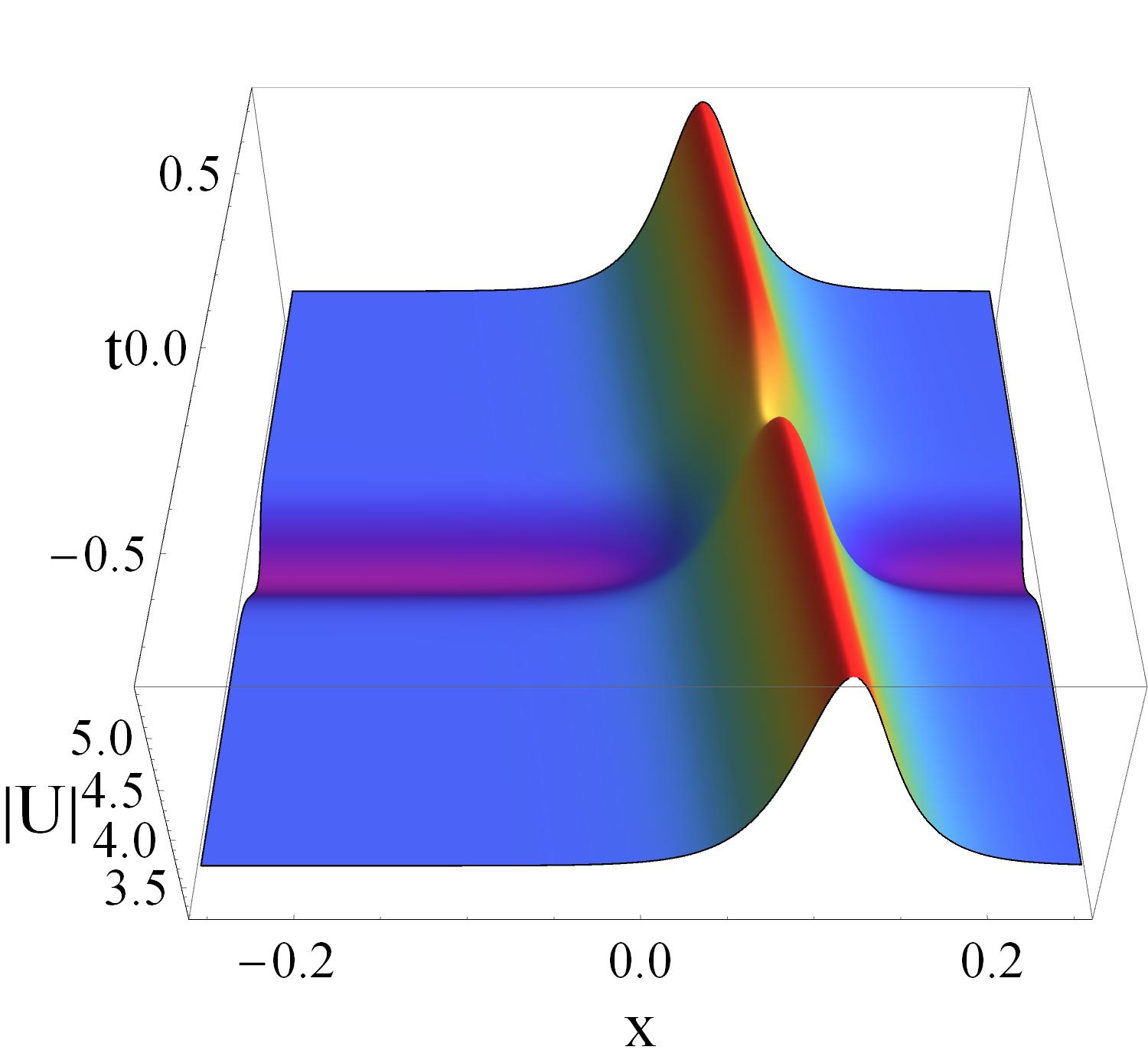}
		\caption{Anti-Dark 1SS}
		\label{Fig3a}
	\end{subfigure}
	\begin{subfigure}{.3\textwidth}
		\includegraphics[width=\textwidth]{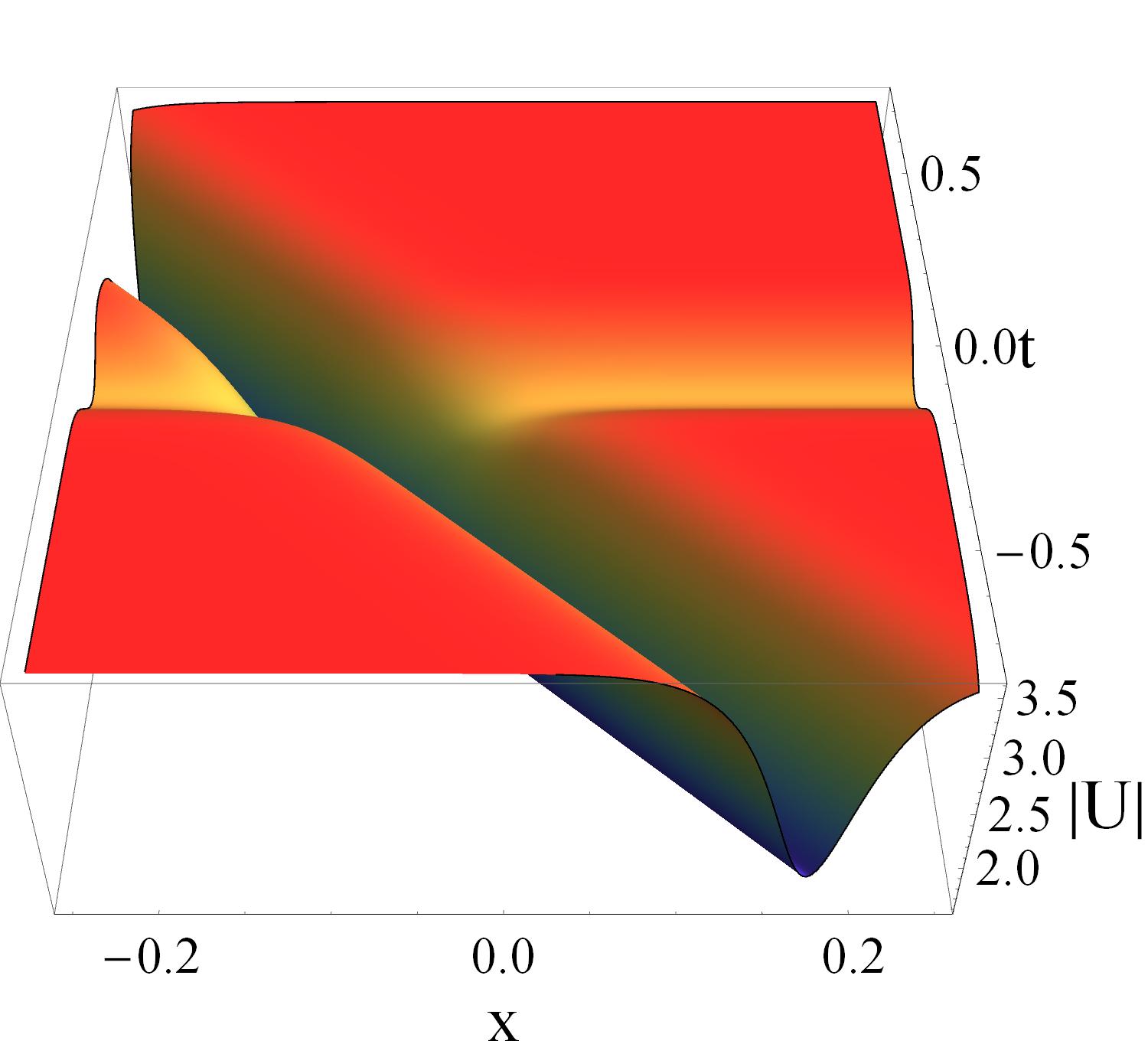}
		\caption{Dark 1SS}
		\label{Fig3b}
	\end{subfigure}
	\begin{subfigure}{.3\textwidth}
		\includegraphics[width=\textwidth]{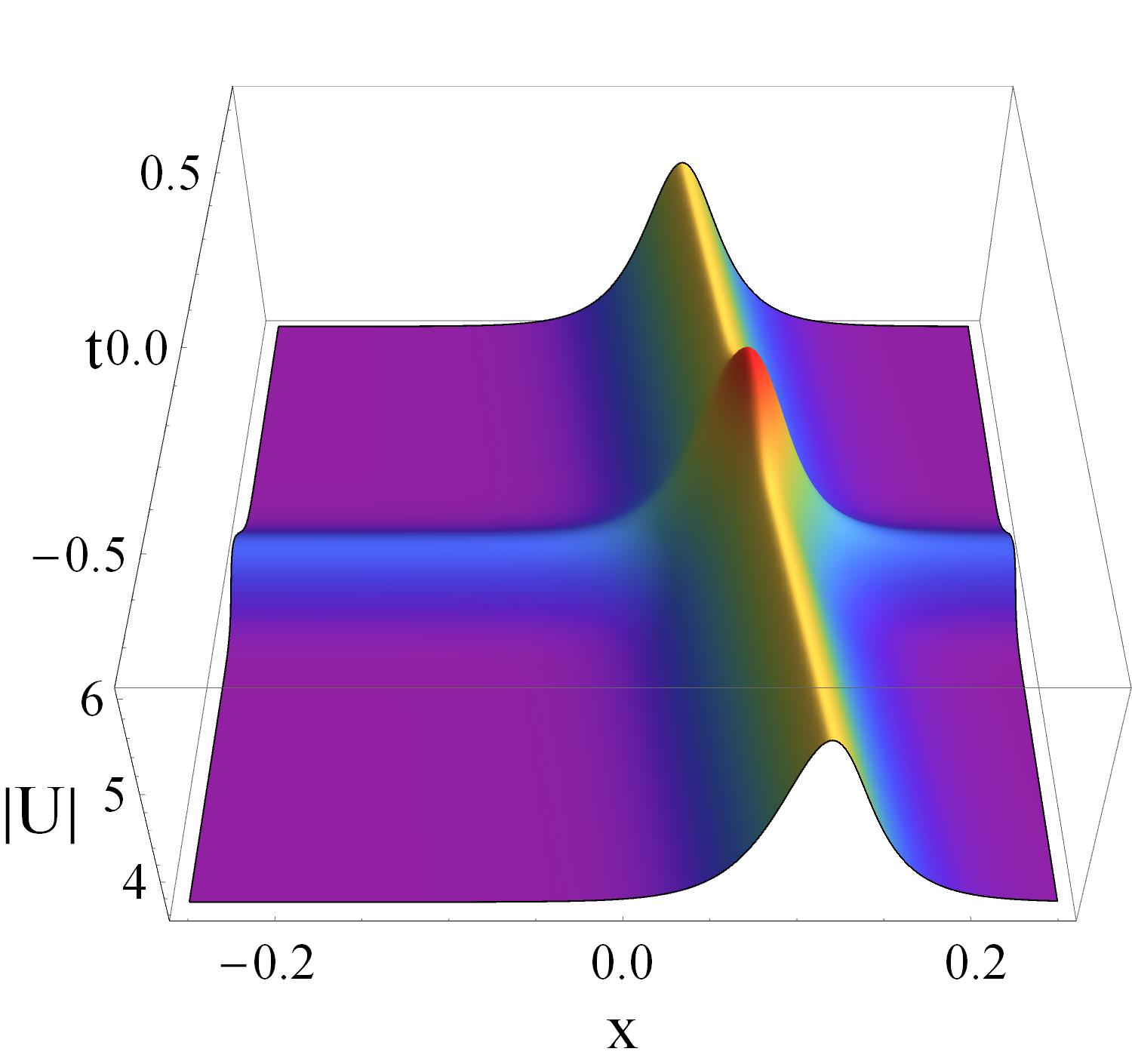}
		\caption{Anti-Dark 1SS}
		\label{Fig3c}
	\end{subfigure}
	\caption{3D plot representation of 1SS under SM scheme: $D(t) = 1$, $R(t) = 1 + \sigma e^{(-k t^2)}$ with $\sigma = 0.25$ and $k = 100$. (a) represents anti-dark soliton and (b) represents dark soliton. (c) represents anti-dark soliton but under SM scheme: $D(t) = 1 + \sigma e^{(-k t^2)}$, $R(t) = 1$.}
	\label{Fig3}
\end{figure}

$D(t)$ and $R(t)$ are linearly independent functions, hence $\Gamma(t) \ne 0$. In this case  the background has a depth for the Gaussian term \say{$\sigma e^{(-k t^2)}$} in $R(t)$ and the gain parameter being non-zero. Figs. (\ref{Fig3a}) - (\ref{Fig3b}) demonstrate this condition. If $D(t)$ and $R(t)$ interchange their values than the background will have a bulge instead of a depth as demonstrated in Fig. (\ref{Fig3c}). Again for the same expressions of $D(t)$ and $R(t)$ (= \say{$1 + \sigma e^{(-k t^2)}$}), $\Gamma(t) = 0$, then for a small value of $\sigma$, the graph will become similar to that of Figs. (\ref{Fig1a}) and (\ref{Fig1b}).

\subsubsection{$D(t) = R(t) = e^{(\sigma t)}\ cos(k t)$ and $\Gamma(\tau)$ vanishes}\label{subsubsec2.1.4}

\begin{figure}
	\centering
	\begin{subfigure}{.45\textwidth}
		\includegraphics[width=\textwidth]{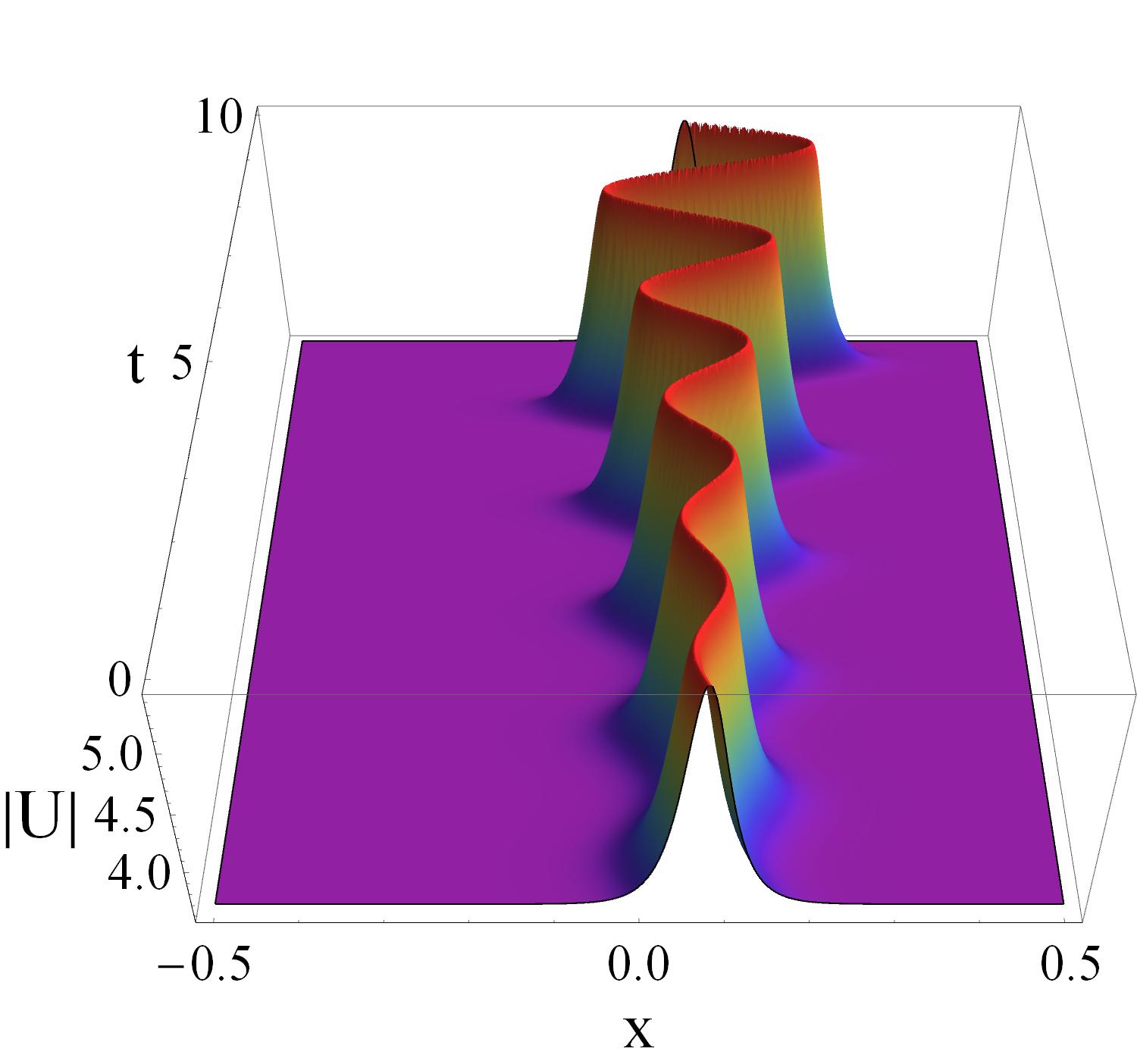}
		\caption{Anti-Dark 1SS}
		\label{Fig4a}
	\end{subfigure}
	\begin{subfigure}{.45\textwidth}
		\includegraphics[width=\textwidth]{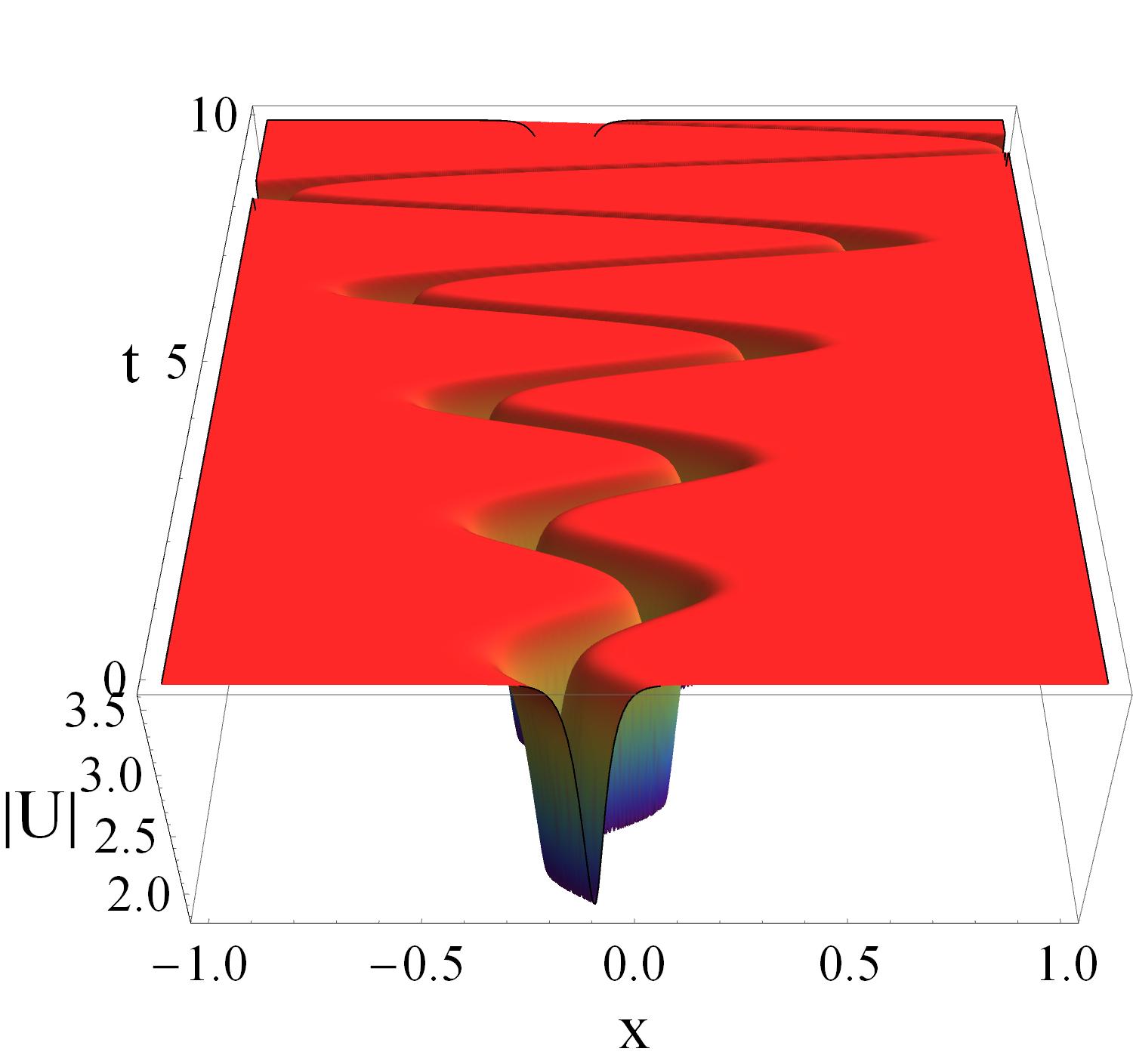}
		\caption{Dark 1SS}
		\label{Fig4b}
	\end{subfigure}
	\caption{3D plot representation of 1SS under SM scheme: $D(t) = R(t) = e^{(\sigma t)}\ cos(k t)$ with $\sigma = 0.25$ and $k = \pi$. (a) represents anti-dark soliton and (b) represents dark soliton.}
	\label{Fig4}
\end{figure}

$D(\tau)$ and $R(\tau)$ are linearly dependent functions, hence $\Gamma(\tau) = 0$. In this case the background is unaffected. 
$D(\tau)$ and $R(\tau)$ are periodic functions and this makes the propagation of the soliton harmonic from mean position ($x = 0$) along the time ($t$) axis. Figs (\ref{Fig4a}) and (\ref{Fig4b}) demonstrate this case. The amplitude remains constant along the propagation of the soliton.

\subsubsection{$D(t) = e^{(\sigma t)}\ cos(k t)$ and $R(t) = cos(k t)$ and $\Gamma(t)$ is non-vanishing}\label{subsubsec2.1.5}

\begin{figure}
	\centering
	\begin{subfigure}{.45\textwidth}
		\includegraphics[width=\textwidth]{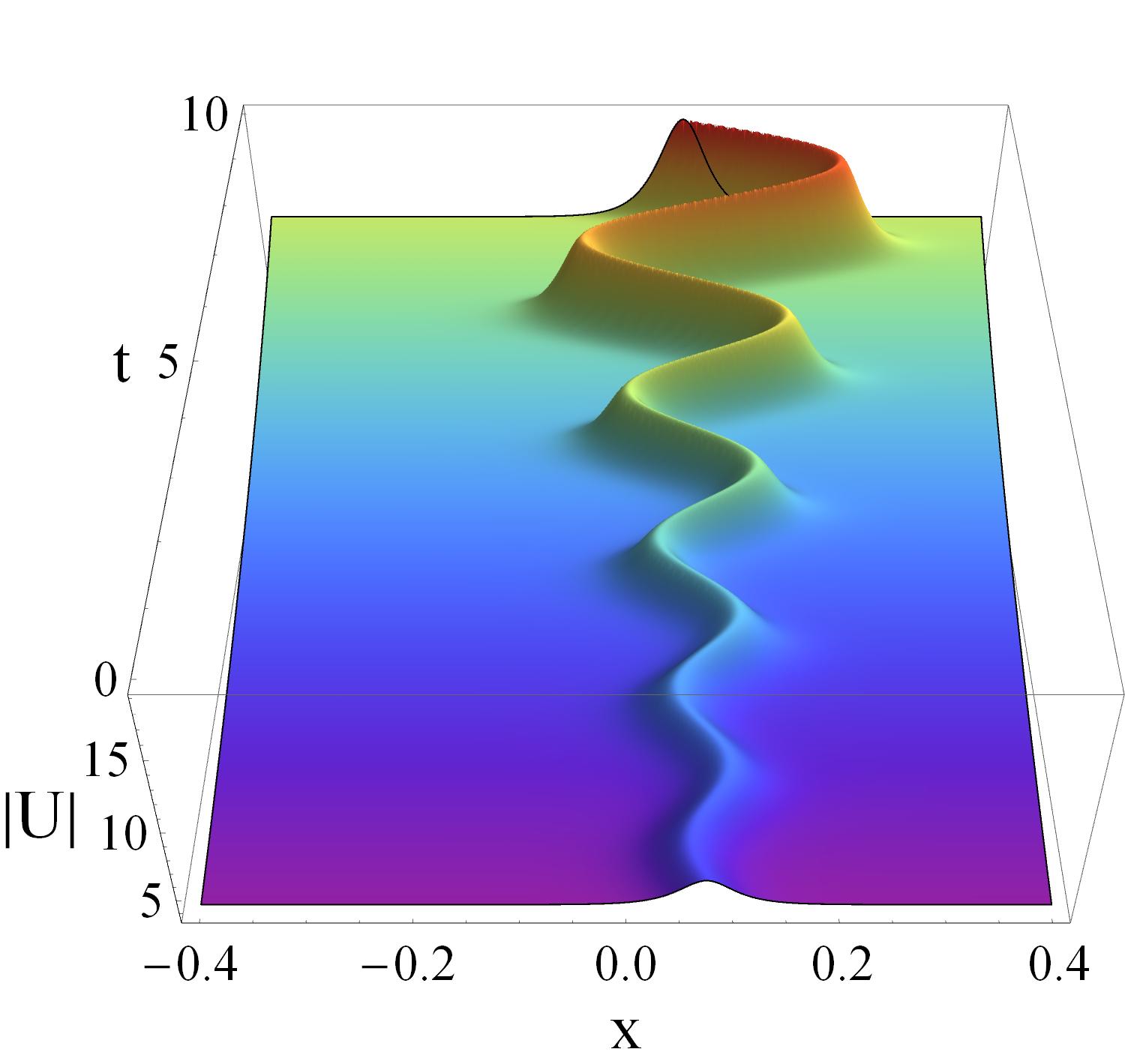}
		\caption{Anti-Dark 1SS}
		\label{Fig5a}
	\end{subfigure}
	\begin{subfigure}{.45\textwidth}
		\includegraphics[width=\textwidth]{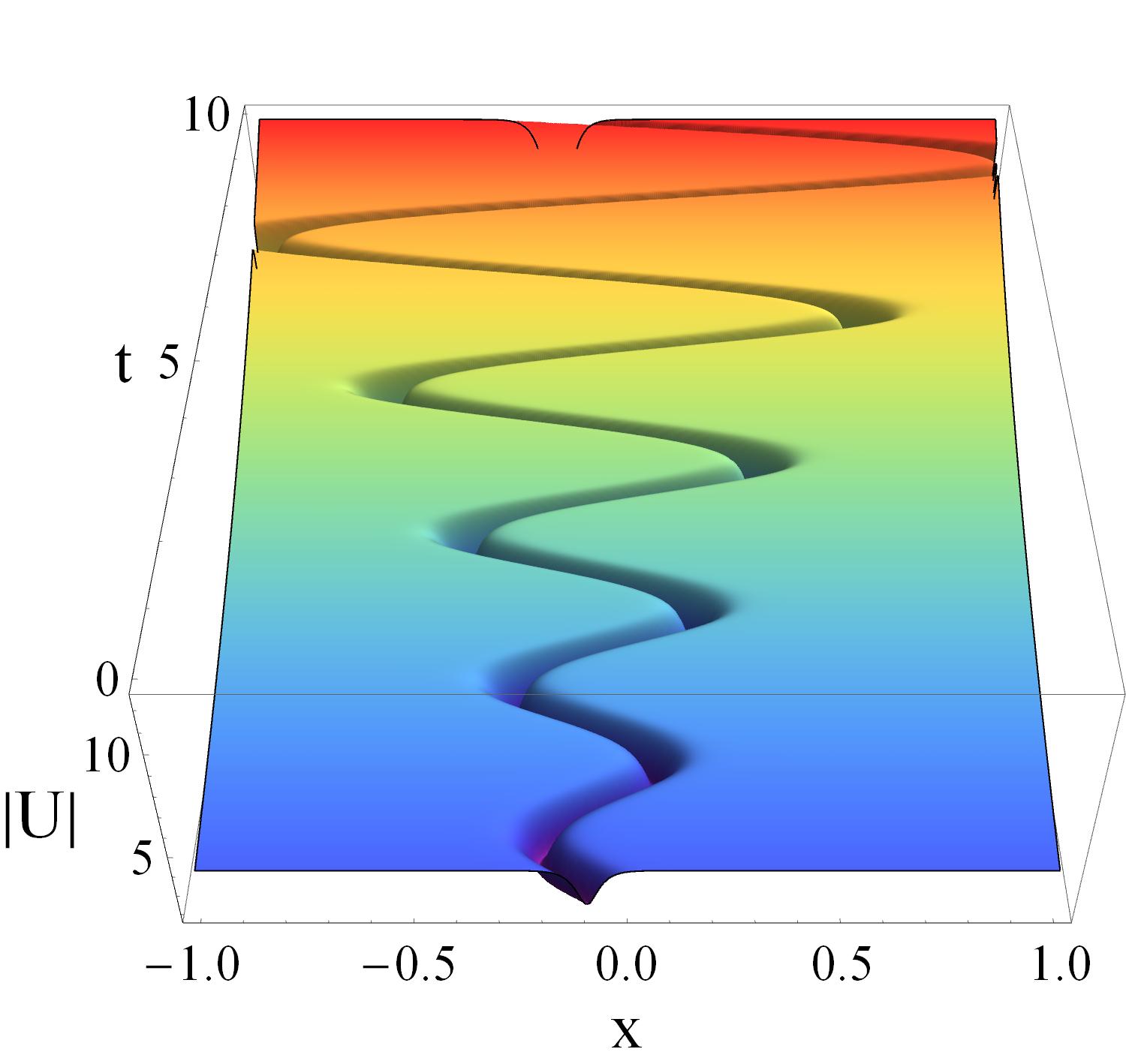}
		\caption{Dark 1SS}
		\label{Fig5b}
	\end{subfigure}
	\caption{3D plot representation of 1SS under SM scheme: $D(t) = e^{(\sigma t)}\ cos(k t)$ and $R(t) = cos(k t)$ with $\sigma = 0.25$ and $k = \pi$. (a) represents anti-dark soliton and (b) represents dark soliton.}
	\label{Fig5}
\end{figure}

$D(\tau)$ and $R(\tau)$ are linearly independent functions and the gain parameter is a non-zero constant, $\Gamma(\tau) = \frac{\sigma}{2}$. In this case the background has a constant gain (or lose) of \say{$\frac{\sigma}{2}$} factor. Just like the previous case (subsection (\ref{subsubsec2.1.4})), here also $D(\tau)$ and $R(\tau)$ are periodic functions and thus the propagation of soliton is harmonic from mean position along the time axis.Figs (\ref{Fig5a}) and (\ref{Fig5b}) demonstrate this case. Amplitude amplification is clearly visible from the graphs unlike the soliton in Figs. (\ref{Fig4}) where the amplitude remains constant. \\

\begin{figure}
	\centering
	\begin{subfigure}{.3\textwidth}
		\includegraphics[width=\textwidth]{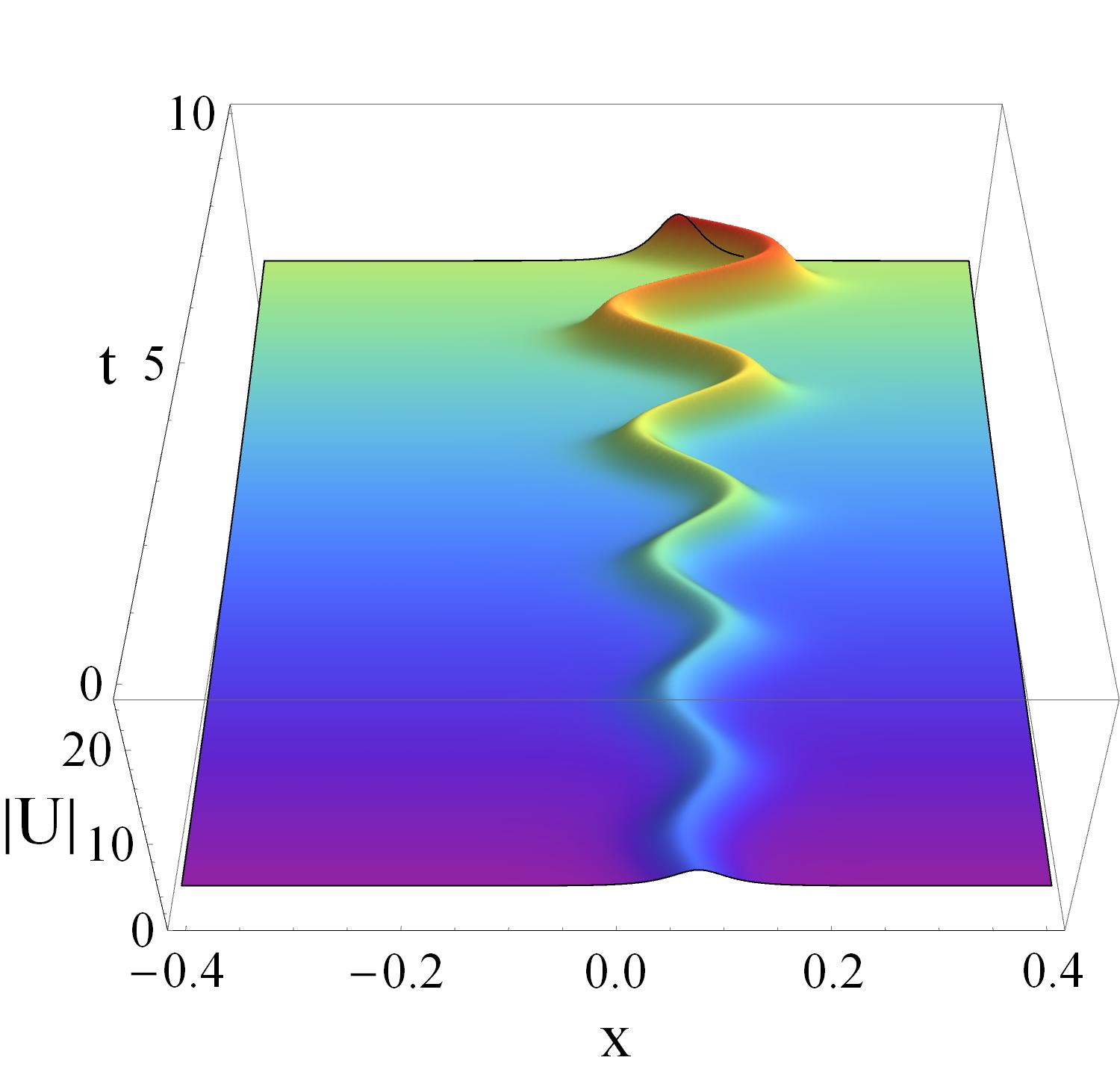}
		\caption{Anti-Dark 1SS}
		\label{Fig6a}
	\end{subfigure}
	\begin{subfigure}{.3\textwidth}
		\includegraphics[width=\textwidth]{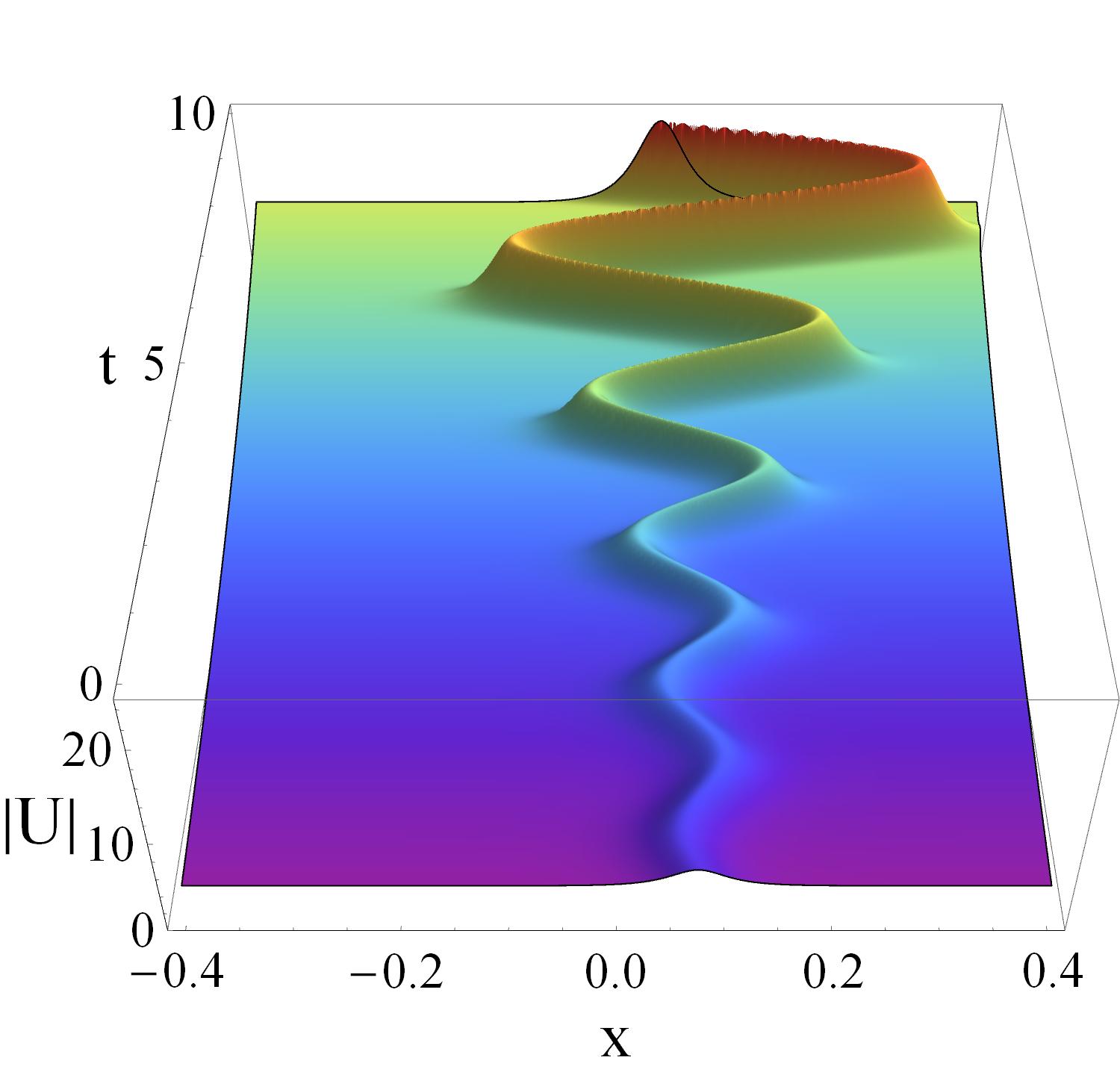}
		\caption{Dark 1SS}
		\label{Fig6b}
	\end{subfigure}
	\begin{subfigure}{.3\textwidth}
		\includegraphics[width=\textwidth]{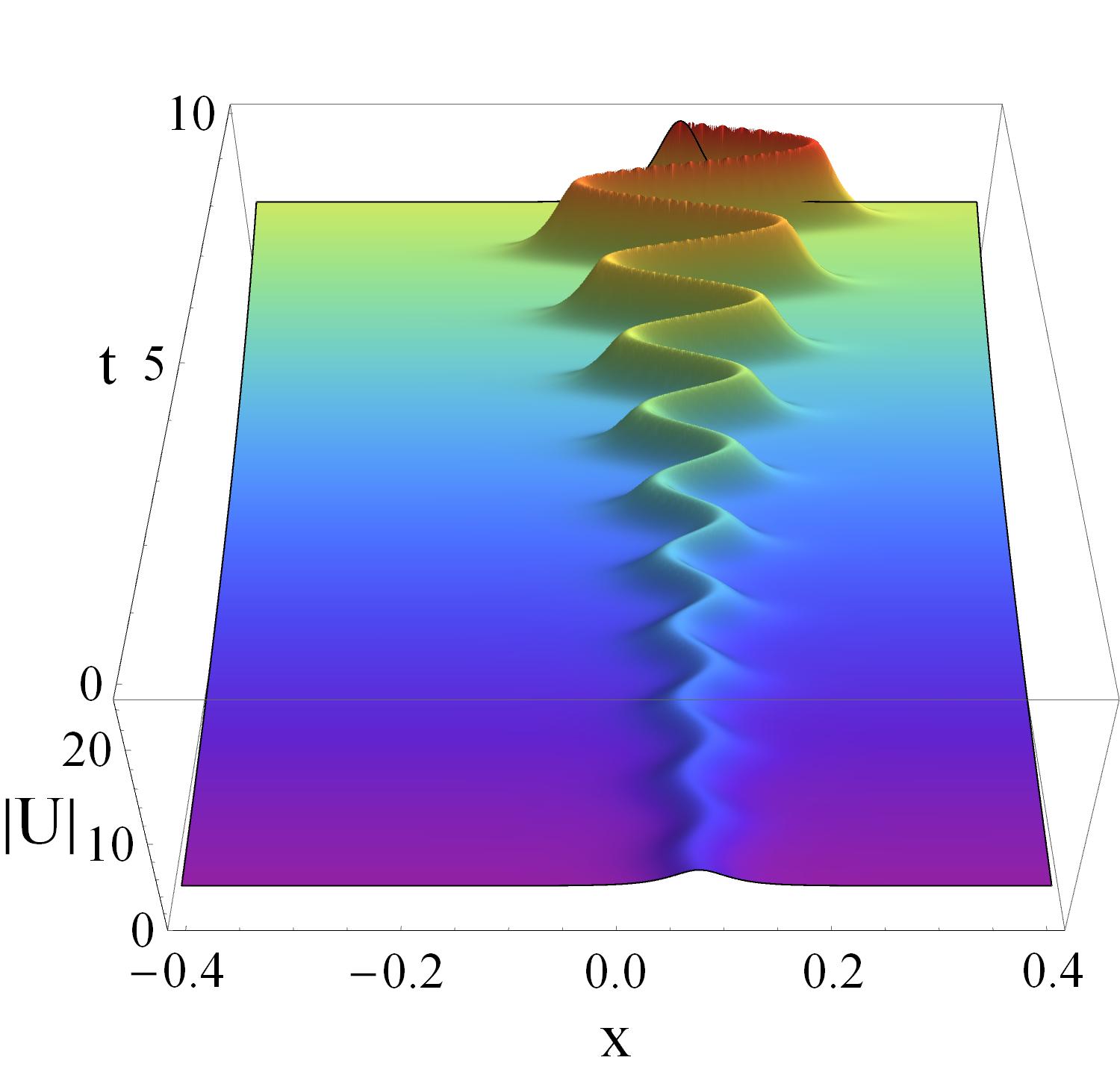}
		\caption{Anti-Dark 1SS}
		\label{Fig6c}
	\end{subfigure}
	\begin{subfigure}{.3\textwidth}
		\includegraphics[width=\textwidth]{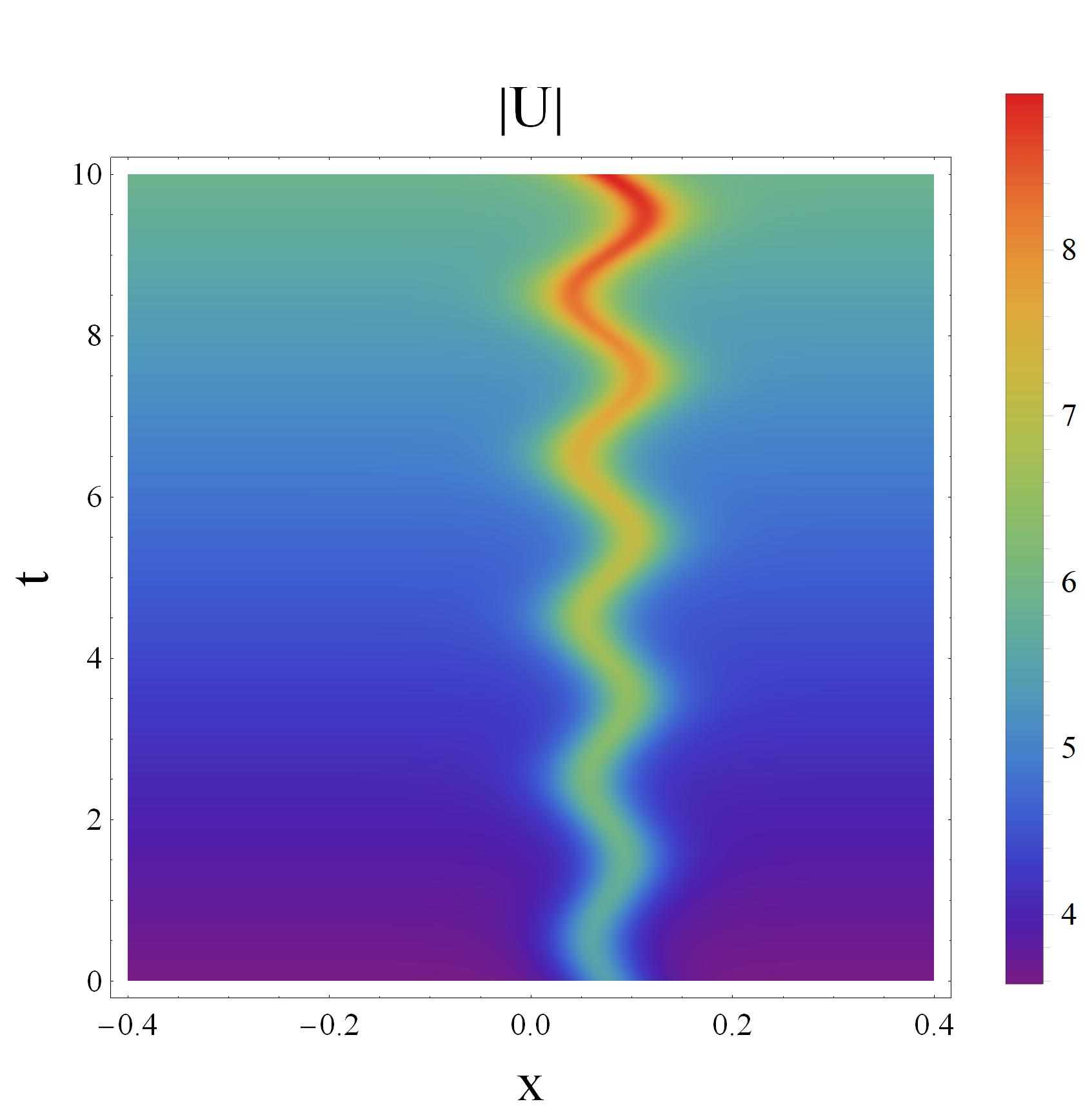}
		\caption{Anti-Dark 1SS}
		\label{Fig6d}
	\end{subfigure}
	\begin{subfigure}{.3\textwidth}
		\includegraphics[width=\textwidth]{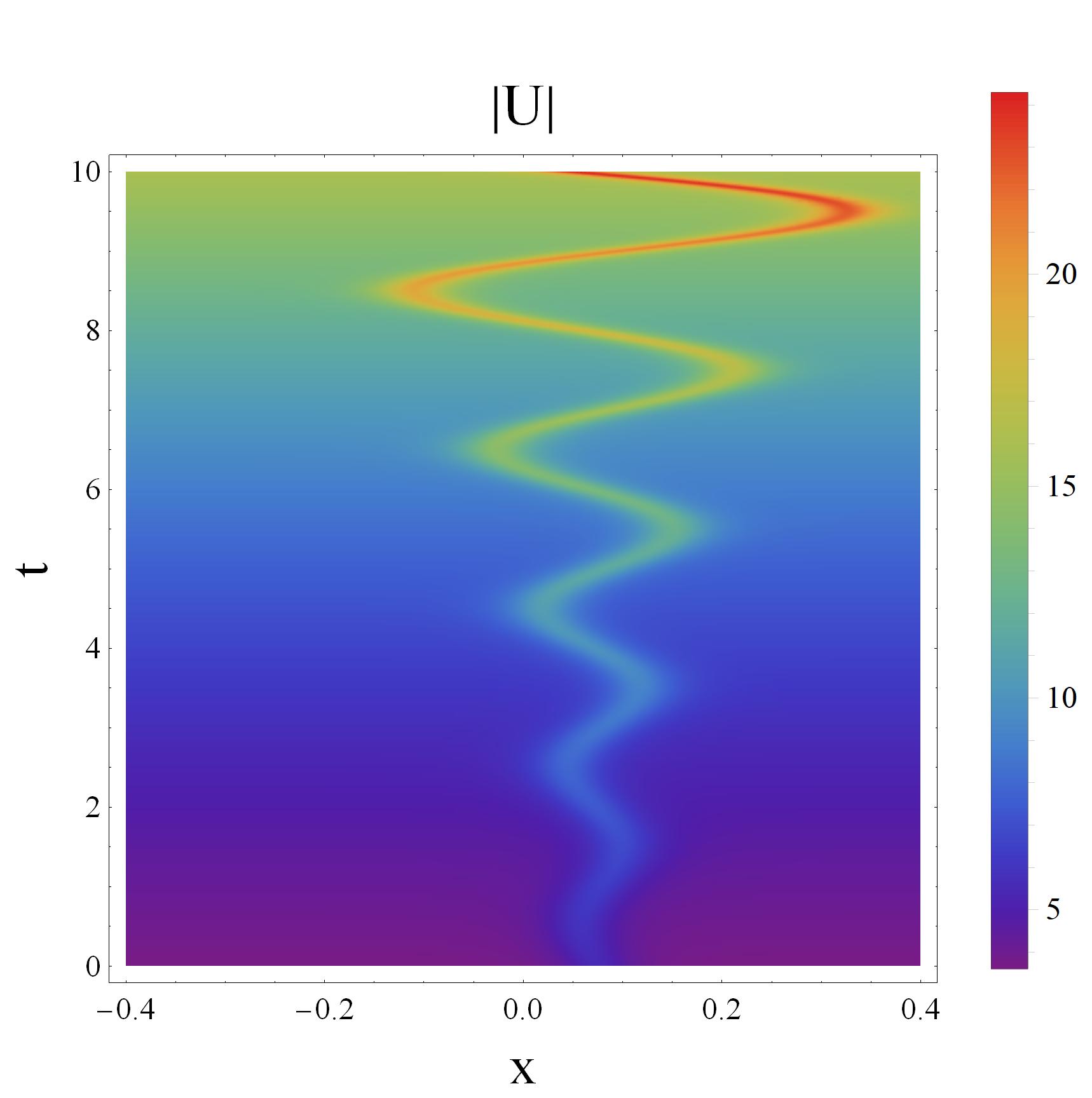}
		\caption{Dark 1SS}
		\label{Fig6e}
	\end{subfigure}
	\begin{subfigure}{.3\textwidth}
		\includegraphics[width=\textwidth]{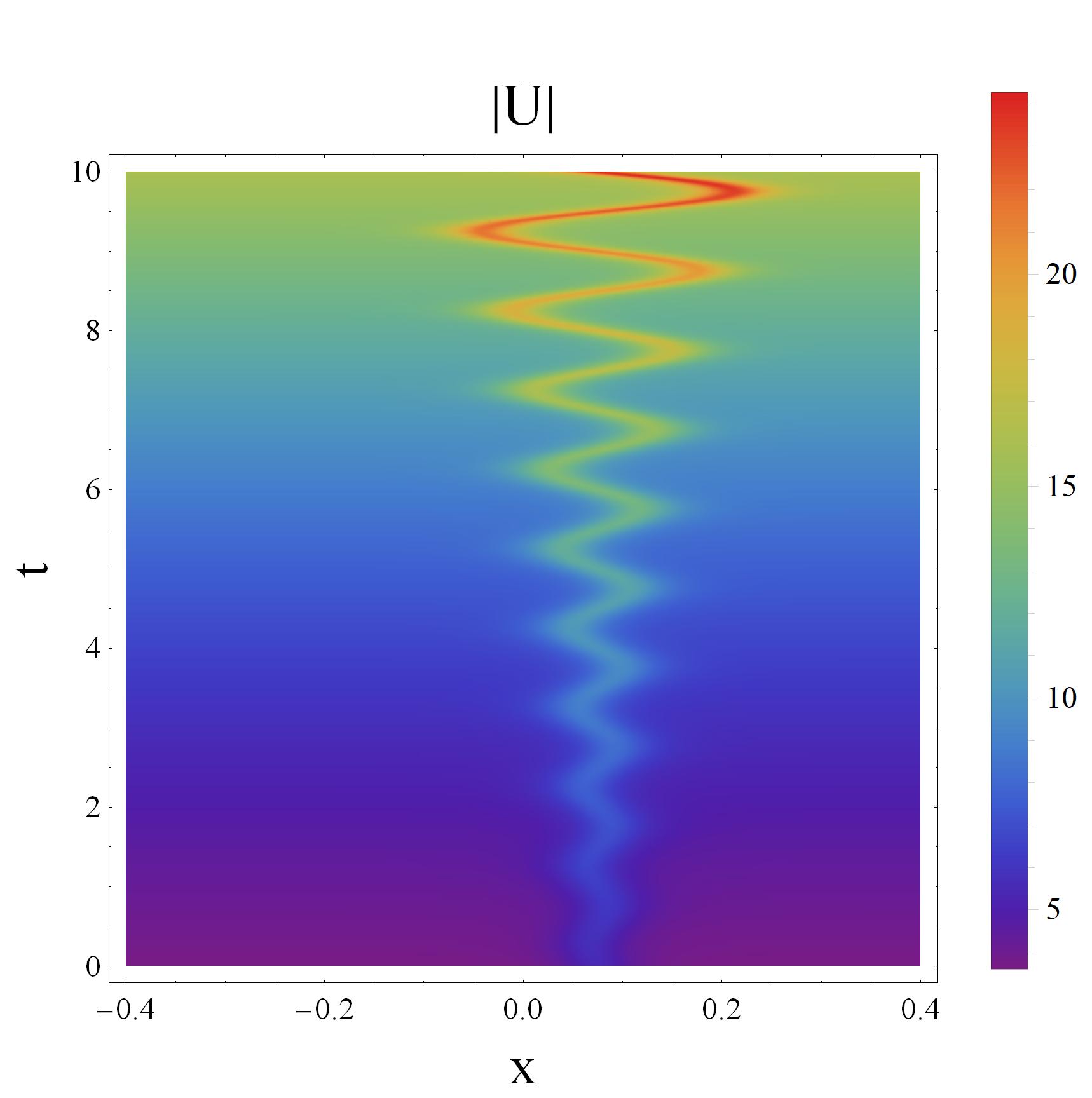}
		\caption{Anti-Dark 1SS}
		\label{Fig6f}
	\end{subfigure}
	\caption{Graphical representation of 1SS under SM scheme: $D(t) = e^{(\sigma t)}\ cos(k t)$ and $R(t) = cos(k t)$. (a), (b) and (c) are 3D plot representation and (d), (e) and (f) are density plot representation. (a) and (d): anti-dark soliton with $\sigma = 0.05$ and $k = 10\pi$, (b) and (e): dark soliton with $\sigma = 0.5$ and $k = 10\pi$ and (c) and (f): anti-dark soliton with $\sigma = 0.5$ and $k = 40\pi$.}
	\label{Fig6}
\end{figure}

Now, exploring the last two cases, i.e. subsections (\ref{subsubsec2.1.4}) and (\ref{subsubsec2.1.5}), we find that $\sigma$ controls the deviation of the soliton from the mean position and $k$ can control the frequency of the soliton and can also keep in check the deviation. In the last subsection (\ref{subsubsec2.1.5}) which is of a non-vanishing constant gain ($\Gamma(t) = \frac{\sigma}{2}$), $\sigma$ can increase (or decrease) the amplitude of the soliton and also affect the background of the system. Comparing Figs. (\ref{Fig6a}) and (\ref{Fig6b}), we can see increasing $\sigma$ from $0.1$ to $0.3$, increases the deviation of soliton from mean position. Again, increasing $k$ from $\pi$ to $2\pi$, increases the frequency of deviating from mean along $t$ axis and also reduces the magnitude of deviation for same time propagation as we can see comparing Figs. (\ref{Fig6c}) with (\ref{Fig6b}). The same are represented as density plots in Figs. (\ref{Fig6d}) - (\ref{Fig6f}) for a more clear view. Keeping in mind the distance required for the soliton to travel, one can fix the gain parameter in such a way that the background change is bearable and the soliton also gain some desirable amplification. With this knowledge while constructing such a SM FLE system, the soliton structure and the propagation behavior can be tuned according to one's requirements. This is a big advantage that the SM scheme provides to a FLE system.

\subsection{Dark and Anti-Dark 2SS}\label{subsec2.2}

In this subsection, we attempt to obtain the 2SS of DLFLE and eventually 2SS of FLE. For this we drop the terms of order greater than or equal to  $\epsilon^5 $ in  $g$, $f$ and $s$. Hence, Eqn. (\ref{bilin0}) for 2SS can be rewritten as

\begin{align}
	\label{sol2}
	u &= \sqrt{\frac{D(\tau)}{R(\tau)}} \frac{g_0 (1 + \epsilon^2 g_2^{(2)} + \epsilon^4 g_4^{(2)})}{1 + \epsilon^2 f_2^{(2)} + \epsilon^4 f_4^{(2)}}\Big|_{\epsilon=1} 
\end{align} 
Let us consider the expressions for $g_2^{(2)}$, $g_4^{(2)}$,$s_2^{(2)}$, $s_4^{(2)}$, $f_2^{(2)}$ and $f_4^{(2)}$ in 2SS are as follows
\begin{align}
	\label{g22}
	g_2^{(2)} &= K_1\ e^{\theta_1 + \theta_1^*} + K_2\ e^{\theta_2 + \theta_2^*} \\
	\label{g4}
	g_4^{(2)} &= K_{12}\ e^{\theta_1 + \theta_1^* + \theta_2 + \theta_2^*} \\
	\label{s22}
	s_2^{(2)} &= M_1\ e^{\theta_1 + \theta_1^*} + M_2\ e^{\theta_2 + \theta_2^*} \\
	\label{s4}
	s_4^{(2)} &= M_{12}\ e^{\theta_1 + \theta_1^* + \theta_2 + \theta_2^*} \\
	\label{f22}
	f_2^{(2)} &= T_1\ e^{\theta_1 + \theta_1^*} + T_2\ e^{\theta_2 + \theta_2^*}\\
	\label{f4}
	f_4^{(2)} &= T_{12}\ e^{\theta_1 + \theta_1^* + \theta_2 + \theta_2^*}
\end{align}
where $\theta_j$ = $p_j\ \xi + \Omega_j\ \tau$. $p_j$, $\Omega_j$, $K_j$, $M_j$, $T_j$, $K_{12}$, $M_{12}$, $T_{12}$ are complex parameters ($j = 1,2$). We assume $p_j$ = $p_{jr} + i\ p_{ji}$ and $T_j$ = $T_{jr} + i\ T_{ji}$ where $p_{jr}$, $p_{ji}$, $T_{jr}$ and $T_{ji}$ are real. Now, substituting the above equations into Eqns. (\ref{BR1}) - (\ref{BR3}), yields the following expressions
\begin{align}
	\label{Omega2}
	\Omega_j &= \frac{h_j}{p_j}\\
	\label{m2}
	M_j &= \frac{T_j^2}{K_j^*}\\
	\label{m12}
	M_{12} &= \frac{T_{12}^2}{K_{12}^*}\\
	\label{k2}
	K_j &= \gamma_j \ T_j^*\\
	\label{k12}
	K_{12} &= \gamma_1\ \gamma_2 \ T_{12}^*\\
	\label{t12}
	T_{12} &= c_{12} \ T_1\ T_2
\end{align}
the expressions for $\gamma_j$'s and $h_j$'s are similar to Eqns. (\ref{gamma}) and (\ref{h}) respectively as
\begin{align}
	\label{gamma2}
	\gamma_j &= \frac{(p_j + p_j^*) T_j - i (T_j - T_j^*) \kappa}{(p_j + p_j^*) T_j^* - i (T_j - T_j^*) \kappa}\\
	\label{h2}
	h_j &= -\frac{|p_j|^2\ (-1 + \gamma_j)^2\ \ \kappa \rho^2}{(p_j + p_j^*)^2 \gamma_j}
\end{align}
and $c_{12}$ is real, expressed as
\begin{align}
	c_{12} &= \frac{(p_{1r}\ T_{1r}\ T_{2i} - 
		p_{2r}\ T_{1i}\ T_{2r})^2 + (p_{1r} - p_{2r})^2\ T_{1i}^2\ T_{2i}^2}{(p_{1r}\ T_{1r}\ T_{2i} - 
		p_{2r}\ T_{1i}\ T_{2r})^2 + (p_{1r} + p_{2r})^2\ T_{1i}^2\ T_{2i}^2}
\end{align}
the 2SS system is under two constraints (for $j = 1, 2$)
\begin{align}
	\label{cons2}
	p_{jr}^2 \ |T_j|^2 + 2 p_{jr} \ \kappa \ (1 + \kappa \rho^2)\ T_{jr} \ T_{ji} + \kappa^2 \ (1 + \kappa \rho^2)\ T_{ji}^2 &= 0
\end{align}
for $T_{jr}$'s and $T_{ji}$'s to be real, there must be restriction of $\ |p_{jr}|\ \le\ \sqrt{ \kappa^3 \rho^2\ (1 + \kappa \rho^2)}$. The amplitude ($A_j$) of the jth soliton is given as
\begin{align}
	\label{A2}
	A_j &= \rho\ \sqrt{\frac{D(\tau)}{R(\tau)}}\ \Big|\frac{T_j + \gamma_j\  |T_j|}{T_j + |T_j|}\Big|
\end{align}
this is similar to Eqn. (\ref{Amp1}), and thus we infer that the nature of the jth soliton to be dark or anti-dark is dependent upon the same criteria of $\kappa$, $p_{jr}$, $T_{ji}$ as that in the 1SS case.

\subsubsection{Asymptotic analysis of 2SS}\label{subsubsec2.2.2}

Soliton has a unique characteristics which is that soliton upon interaction with other soliton, retains all of its characteristics like shape, size, etc. with a shift in its phase. To study the amplitudes and the phase shift of the individual solitons of 2SS, we are going to perform asymptotic analysis. The individual solitons of 2SS when asymptotically apart, are essentially two separated 1SS. In the limit of before interaction $\tau \to -\infty$, fixing $\theta_1$ to a finite value implies $|e^{\theta_1}|$ is finite and $|e^{\theta_2}| \to \infty$. In this case, Eqn. (\ref{sol2}) takes the form
\begin{align}
	\label{ua1kt}
	u_{1-} &=  \rho\ e^{i\ (\kappa \xi + \omega(\tau))}\ \sqrt{\frac{D(\tau)}{R(\tau)}}\ \frac{K_2\ (1 + c_{12}\ K_1\ e^{\theta_1 + \theta_1^*})}{T_2\ (1 + c_{12}\ T_1\ e^{\theta_1 + \theta_1^*})}\\
	\label{ua1}
	\Rightarrow u_{1-} &=  \rho\ e^{i\ (\kappa \xi + \omega(\tau))}\ \sqrt{\frac{D(\tau)}{R(\tau)}}\ e^{i \phi_{1}}\ \frac{(1 + K_1^\prime\ e^{\theta_1 + \theta_1^*})}{(1 + T_1^\prime\ e^{\theta_1 + \theta_1^*})}
\end{align}
here $ K_1^\prime = c_{12}\ K_1$ and  $T_1^\prime = c_{12}\ T_1$. Eqn. (\ref{ua1}) is simply an expression for 1SS with an additional phase $\phi_{1}$. The fractional term \say{$\frac{K_2}{T_2}$} in Eqn. (\ref{ua1kt}) has a magnitude of 1 but a phase difference of $\phi_{1}$ which results the phase term \say{$e^{i \phi_{1}}$} in Eqn. (\ref{ua1}). \\

In the limit of after interaction $\tau \to \infty$ and for a fixed finite value of $\theta_1$, we have a finite $|e^{\theta_1}|$ and $|e^{\theta_2}| \to 0$. This time, the corresponding asymptotic form of Eqn. (\ref{sol2}) becomes
\begin{align}
	\label{ua2}
	u_{1+} &=  \rho\ e^{i\ (\kappa \xi + \omega(\tau))}\ \sqrt{\frac{D(\tau)}{R(\tau)}}\ \frac{(1 + K_1\ e^{\theta_1 + \theta_1^*})}{(1 + T_1\ e^{\theta_1 + \theta_1^*})}
\end{align}
we see Eqn. (\ref{ua2}) is also in the form of 1SS just like Eqn. (\ref{ua1}) with a difference in phase. This phase shift $\phi_{1}$ is given as
\begin{align}
	\label{phi1}
	\phi_{1} &= tan^{-1}\Big[-\frac{2 \kappa\ T_{i_2}^2\ (\kappa\ T_{r_2}\ T_{i_2} + p_{r_2} |T_2|^2)}{(\kappa\ (T_{r_2} - T_{i_2})\ T_{i_2} + p_{r_2} |T_2|^2)\ (\kappa\ (T_{r_2} + T_{i_2})\ T_{i_2} + p_{r_2} |T_2|^2)}\Big]
\end{align}

The amplitudes of both the Eqns. (\ref{ua1}) and (\ref{ua2}) are same and can be expressed through Eqn. (\ref{Amp1}). This case represents a purely elastic collision between two solitons since each of them retains all their characteristics with a slight change in phase. We can perform the same analysis by keeping $\theta_2$ finite. This time also, the amplitude before and after interaction remains the same and the phase shift can be represented in a similar way as in Eqn. (\ref{phi1}) but by replacing the components of $p_2$ and $T_2$ with the respective components of $p_1$ and $T_1$. With this, we can infer that our system DLFLE with a non-vanishing background condition supports elastic soliton interaction. \\

\begin{figure}
	\centering
	\begin{subfigure}{.45\textwidth}
		\includegraphics[width=\textwidth]{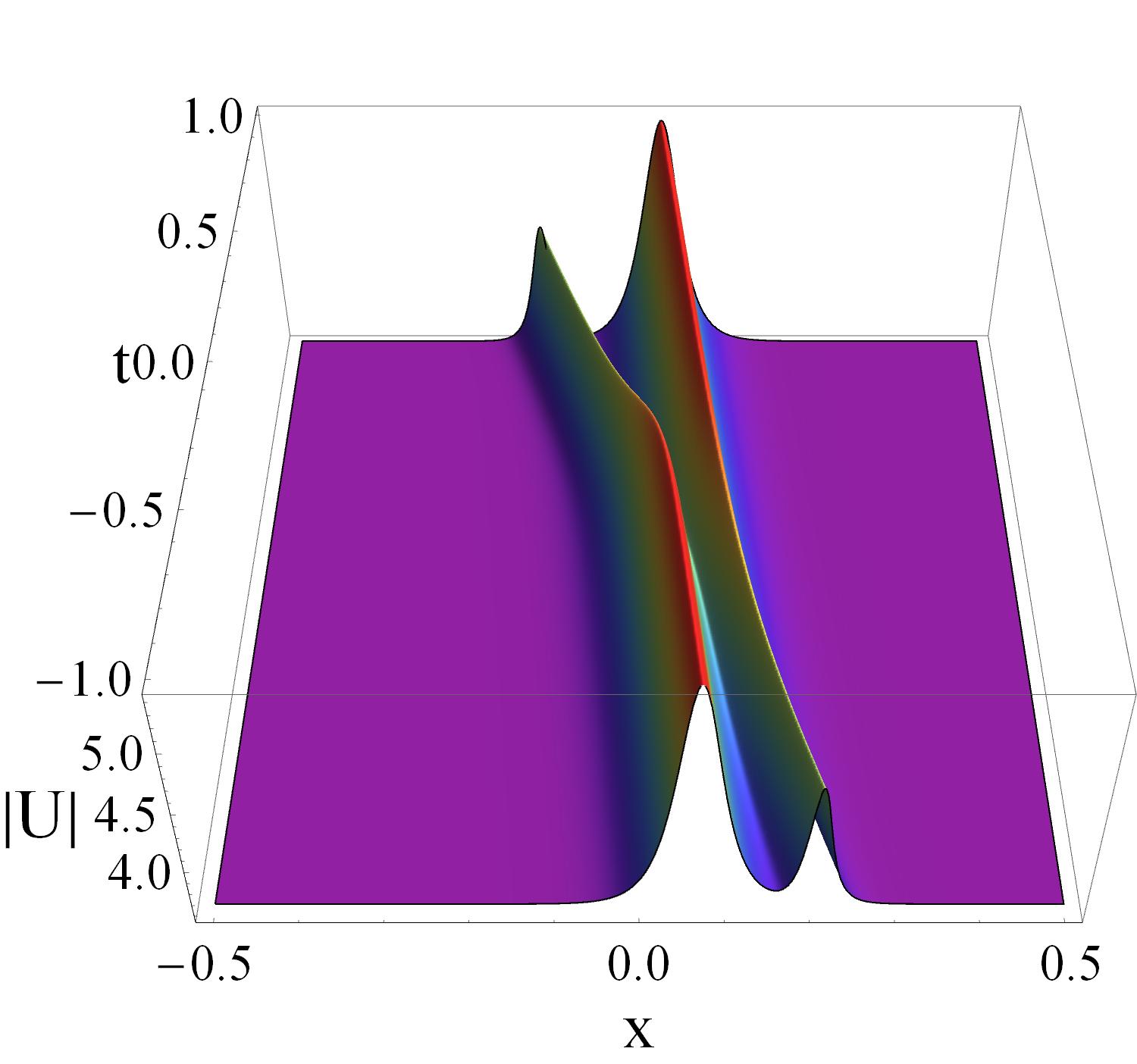}
		\caption{Anti-dark-Anti-dark 2SS}
		\label{Fig7a}
	\end{subfigure}
	\begin{subfigure}{.45\textwidth}
		\includegraphics[width=\textwidth]{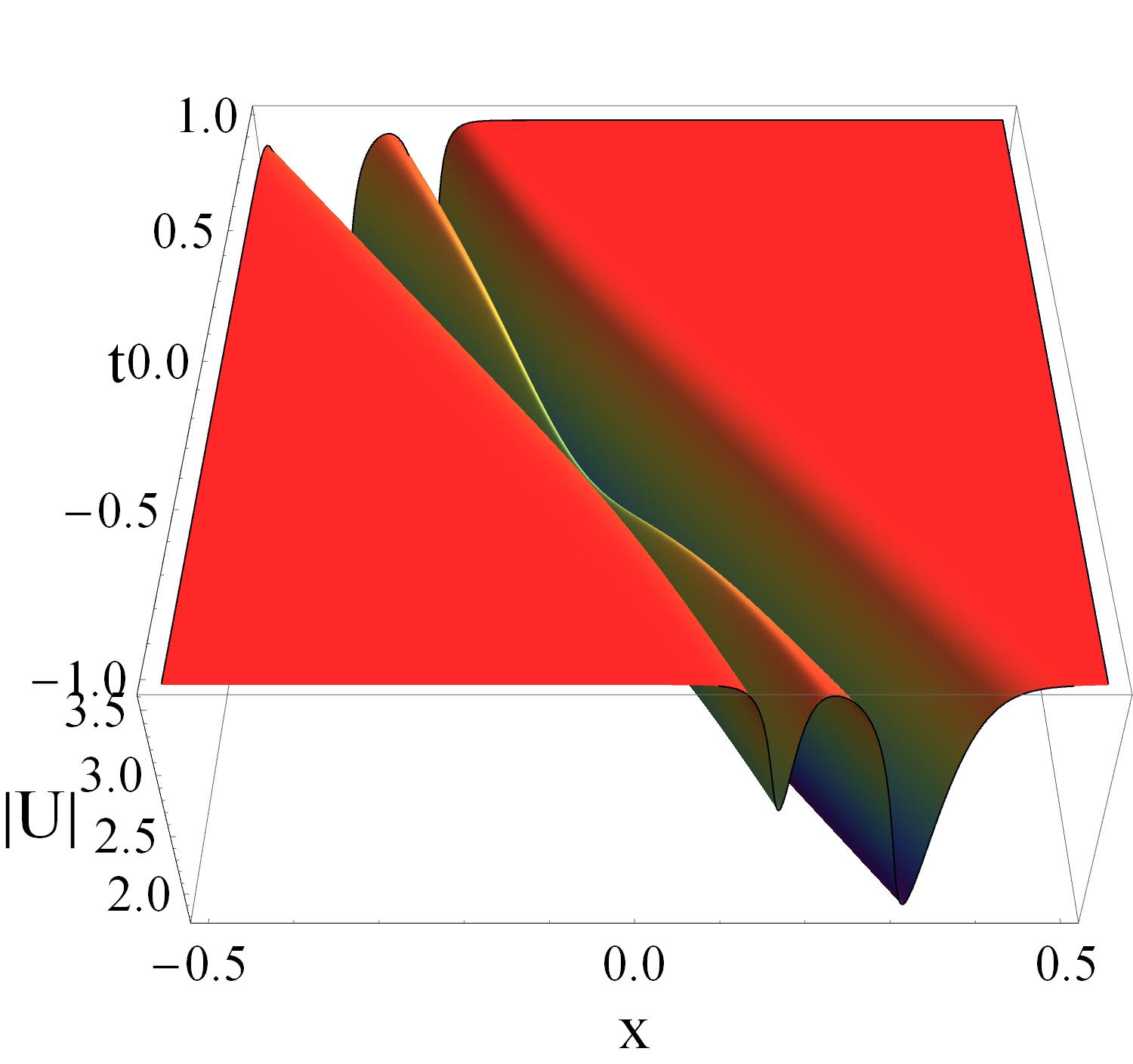}
		\caption{Dark-Dark 2SS}
		\label{Fig7b}
	\end{subfigure}
	\begin{subfigure}{.45\textwidth}
		\includegraphics[width=\textwidth]{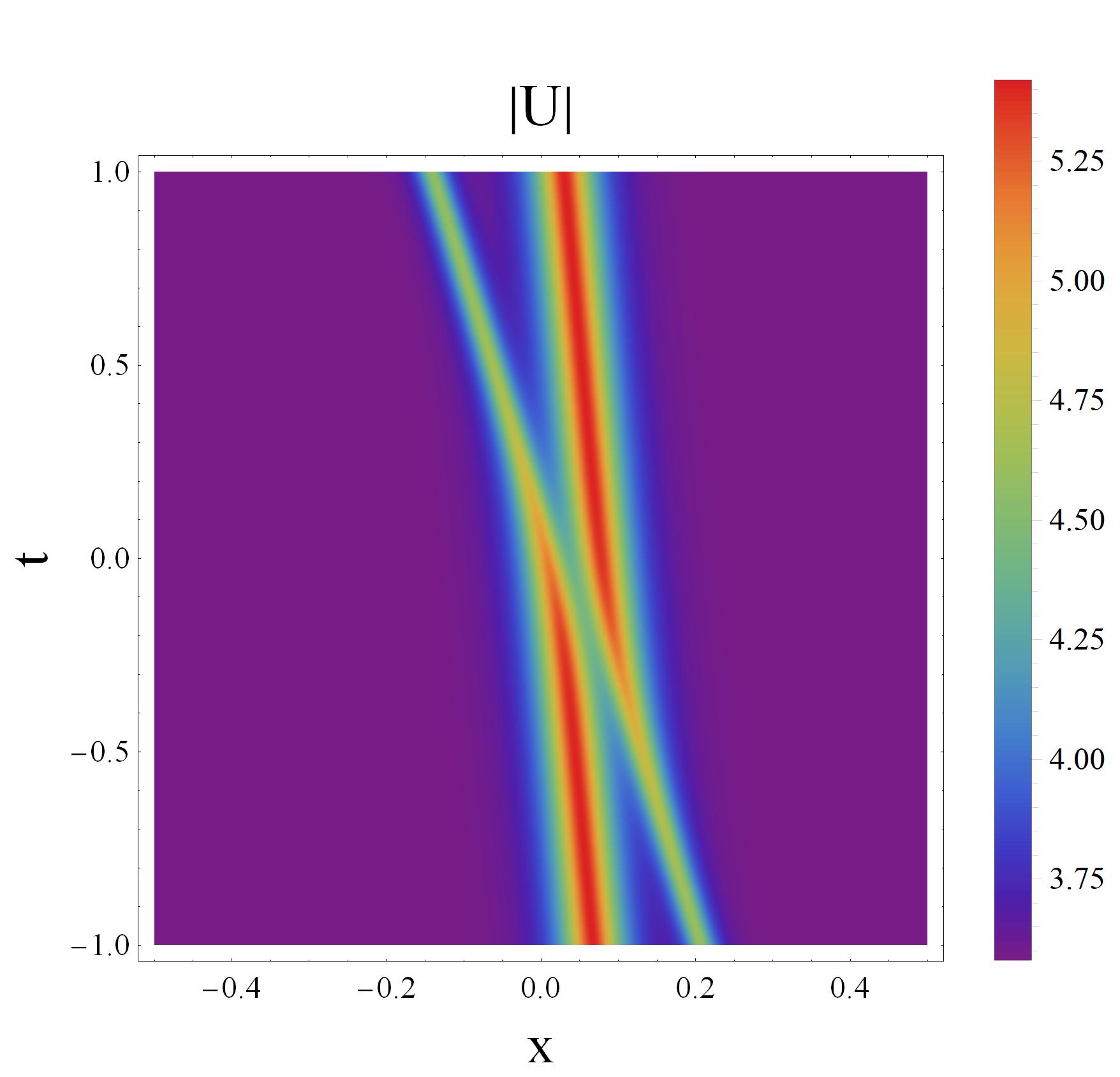}
		\caption{Anti-dark-Anti-dark 2SS}
		\label{Fig7c}
	\end{subfigure}
	\begin{subfigure}{.45\textwidth}
		\includegraphics[width=\textwidth]{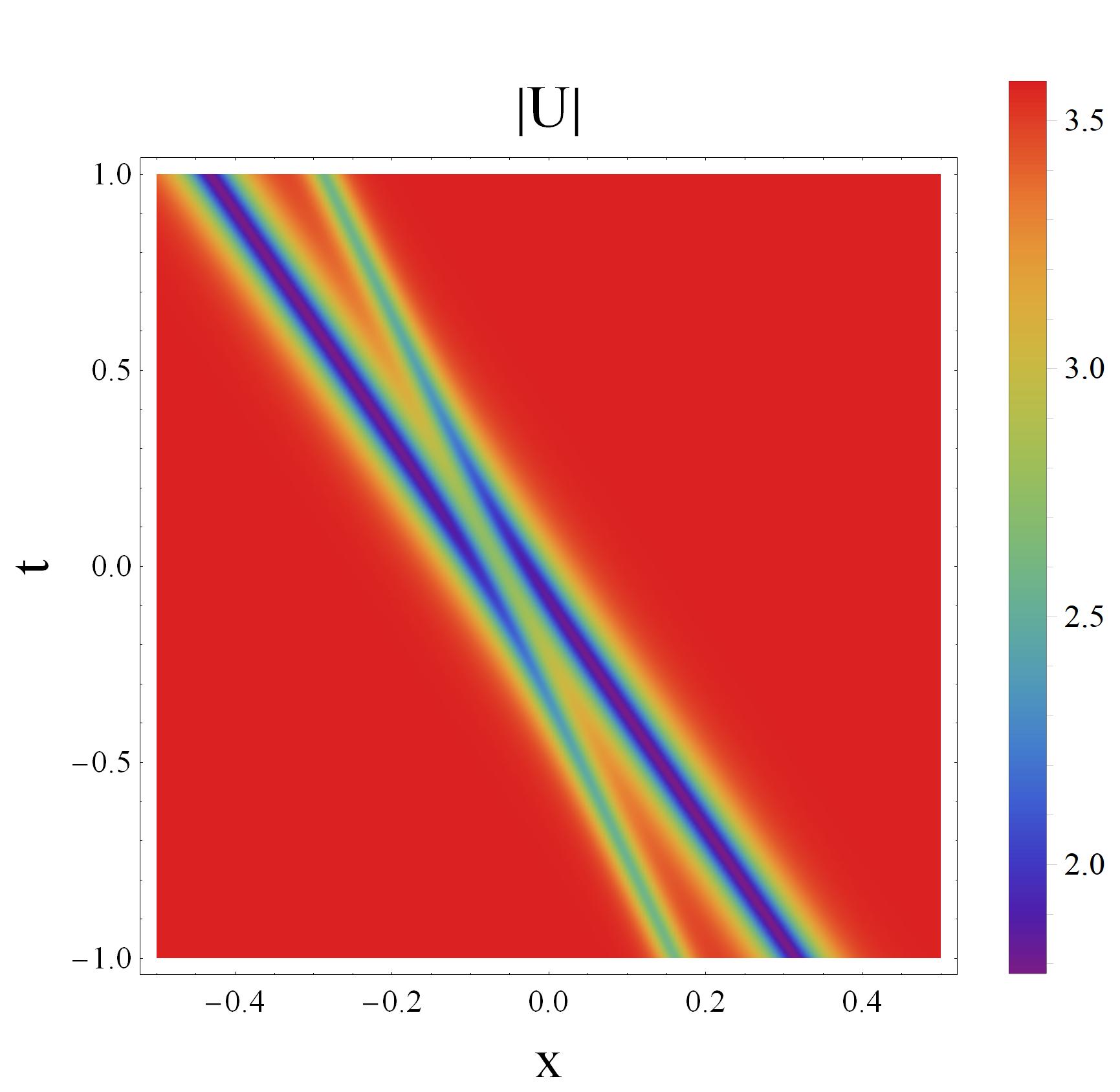}
		\caption{Dark-Dark 2SS}
		\label{Fig7d}
	\end{subfigure}
	\caption{Graphical representation of 2SS under SM scheme: $D(t) = R(t) = 1$. (a) and (b) are 3D plot representation and (c) and (d) are density plot representation. (a) and (c) represent anti-dark-anti-dark soliton with $p_{1r} = -10$, $p_{2r} = -20$, $\kappa = 5$, $T_{1r} = -6$, $T_{2r} = -9$ and (b) and (d) represent dark-dark soliton with $p_{1r} = 10$, $p_{2r} = 20$, $\kappa = -5$, $T_{1r} = 6$, $T_{2r} = 9$. $T_{1i}$ and $T_{2i}$ are calculated using Eqn. (\ref{cons2}). In all the four graphs, we fix $\rho = 2$, $m = 0.2$, $n = 4$, $b = 1$, $a_2 = \frac{1}{n}$, $a_1 = m\ a_2$.}
	\label{Fig7}
\end{figure}

Finally, just like in 1SS case, here also we get the 2SS of FLE by putting Eqn. (\ref{sol2}) in Eqn. (\ref{GT}). Figs. (\ref{Fig7a}) - (\ref{Fig7d}) demonstrate 2SS under SM scheme: $D(t) = R(t) = 1$ and hence $\Gamma(t) = 0$. This implies that the background is unaffected. We can observe from the two density plots (Figs. (\ref{Fig7c}) - (\ref{Fig7d})) that the individual solitons even upon interaction with each other, propagates with the same characteristics as was before interaction, but only a shift occurs in their respective phases. \\

\begin{figure}
	\centering
	\begin{subfigure}{.3\textwidth}
		\includegraphics[width=\textwidth]{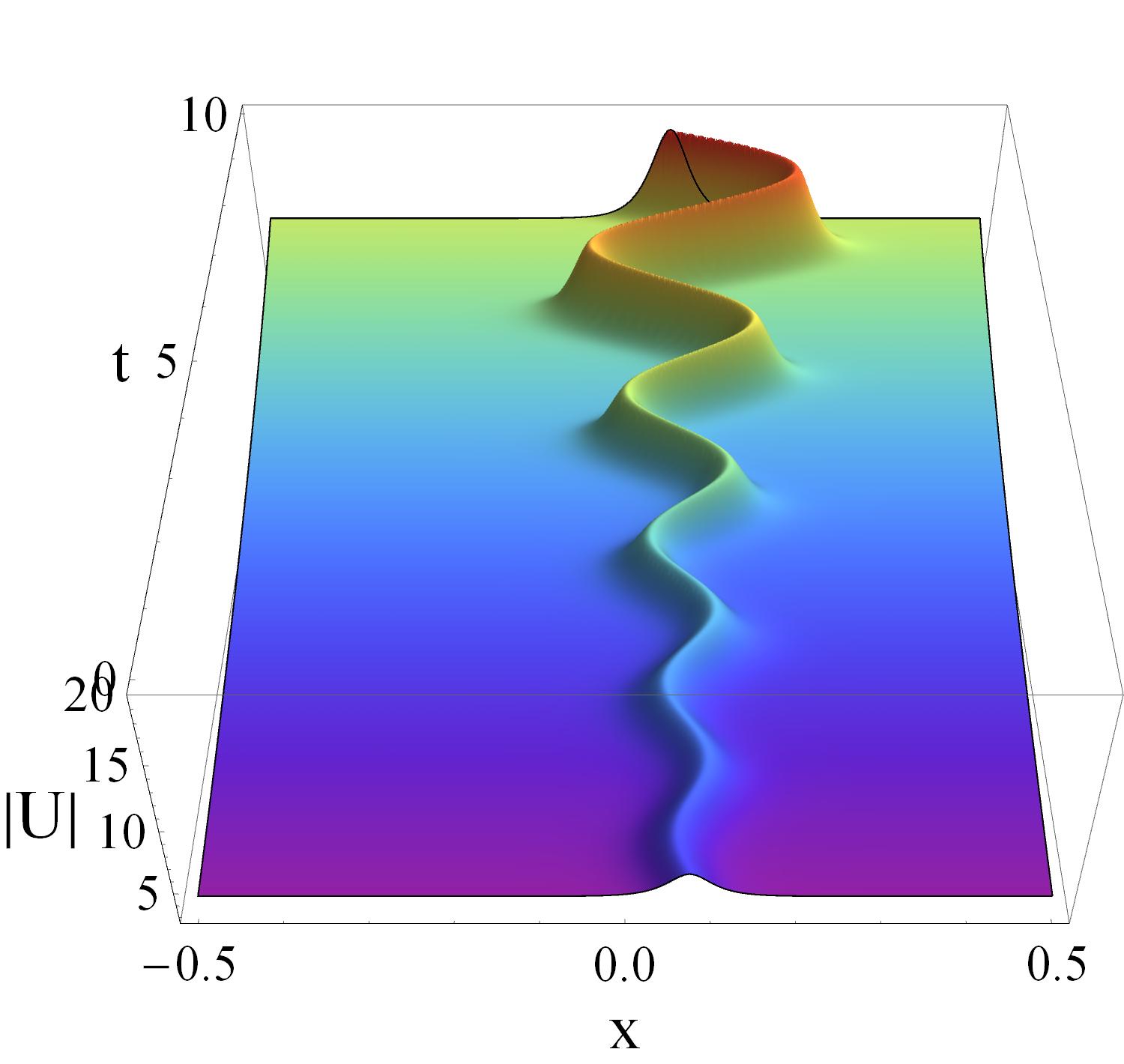}
		\caption{Anti-Dark 1SS}
		\label{Fig8a}
	\end{subfigure}
	\begin{subfigure}{.3\textwidth}
		\includegraphics[width=\textwidth]{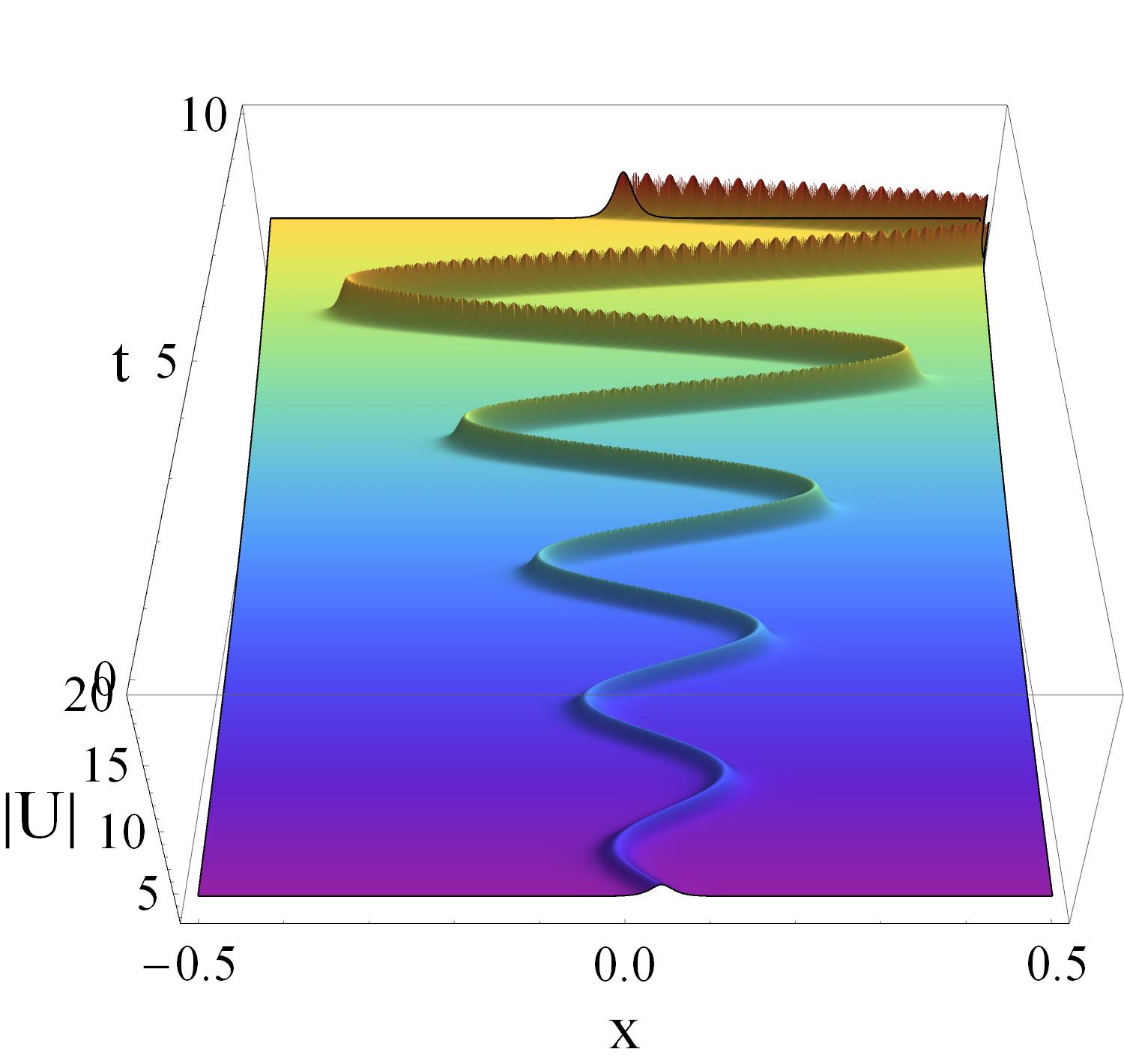}
		\caption{Anti-Dark 1SS}
		\label{Fig8b}
	\end{subfigure}
	\begin{subfigure}{.3\textwidth}
		\includegraphics[width=\textwidth]{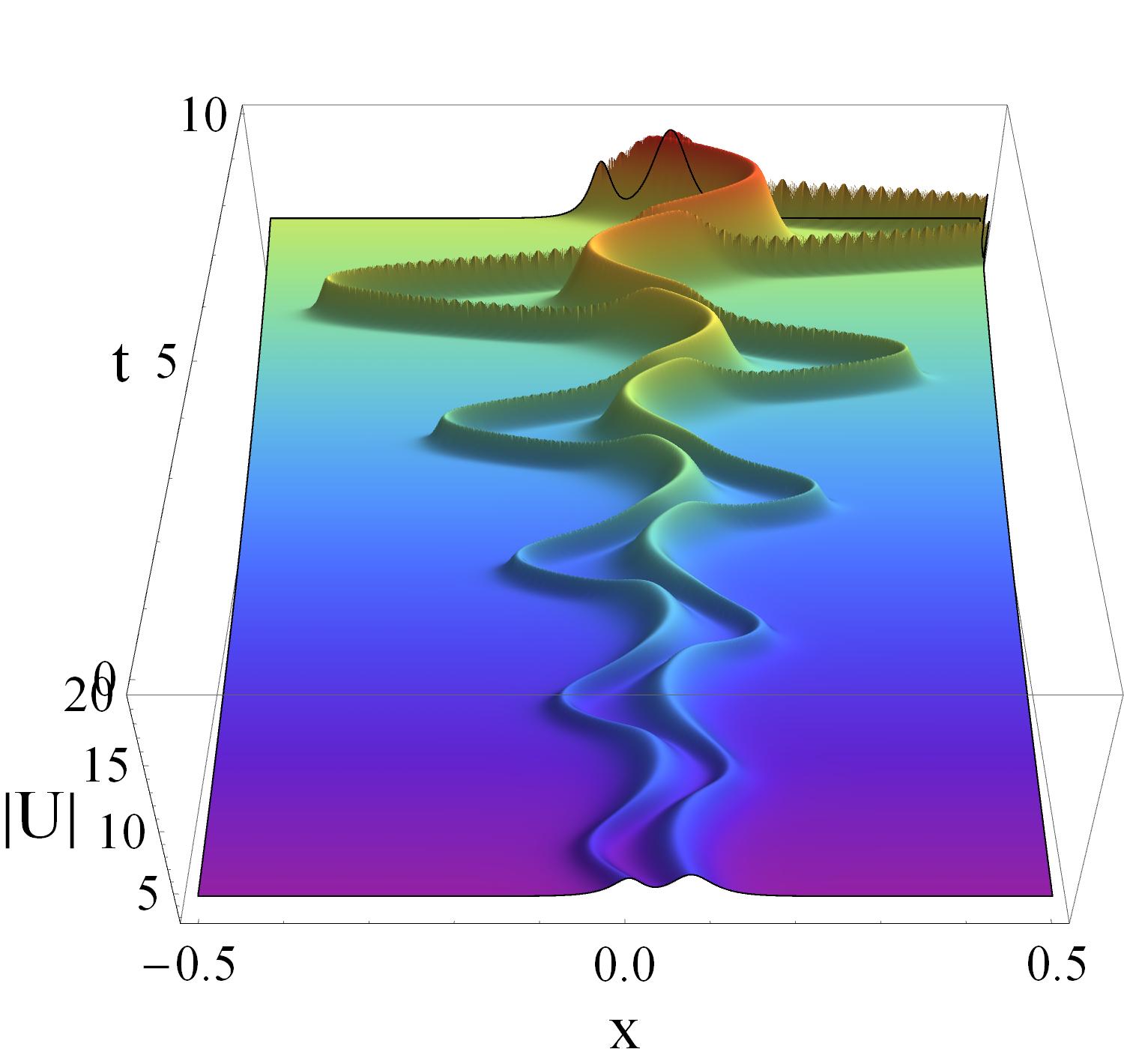}
		\caption{Anti-Dark 2SS}
		\label{Fig8c}
	\end{subfigure}
	\begin{subfigure}{.3\textwidth}
		\includegraphics[width=\textwidth]{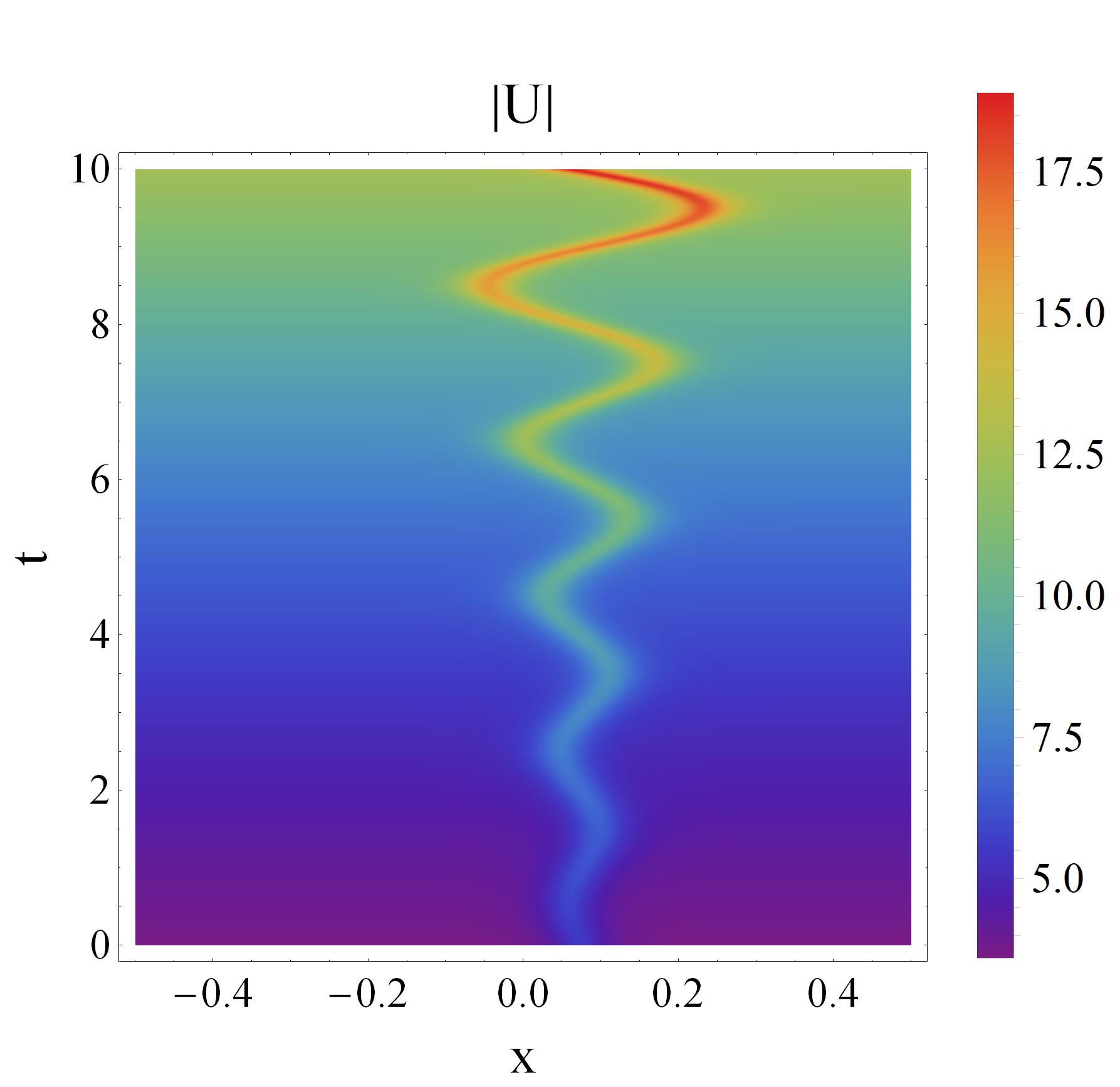}
		\caption{Anti-Dark 1SS}
		\label{Fig8d}
	\end{subfigure}
	\begin{subfigure}{.3\textwidth}
		\includegraphics[width=\textwidth]{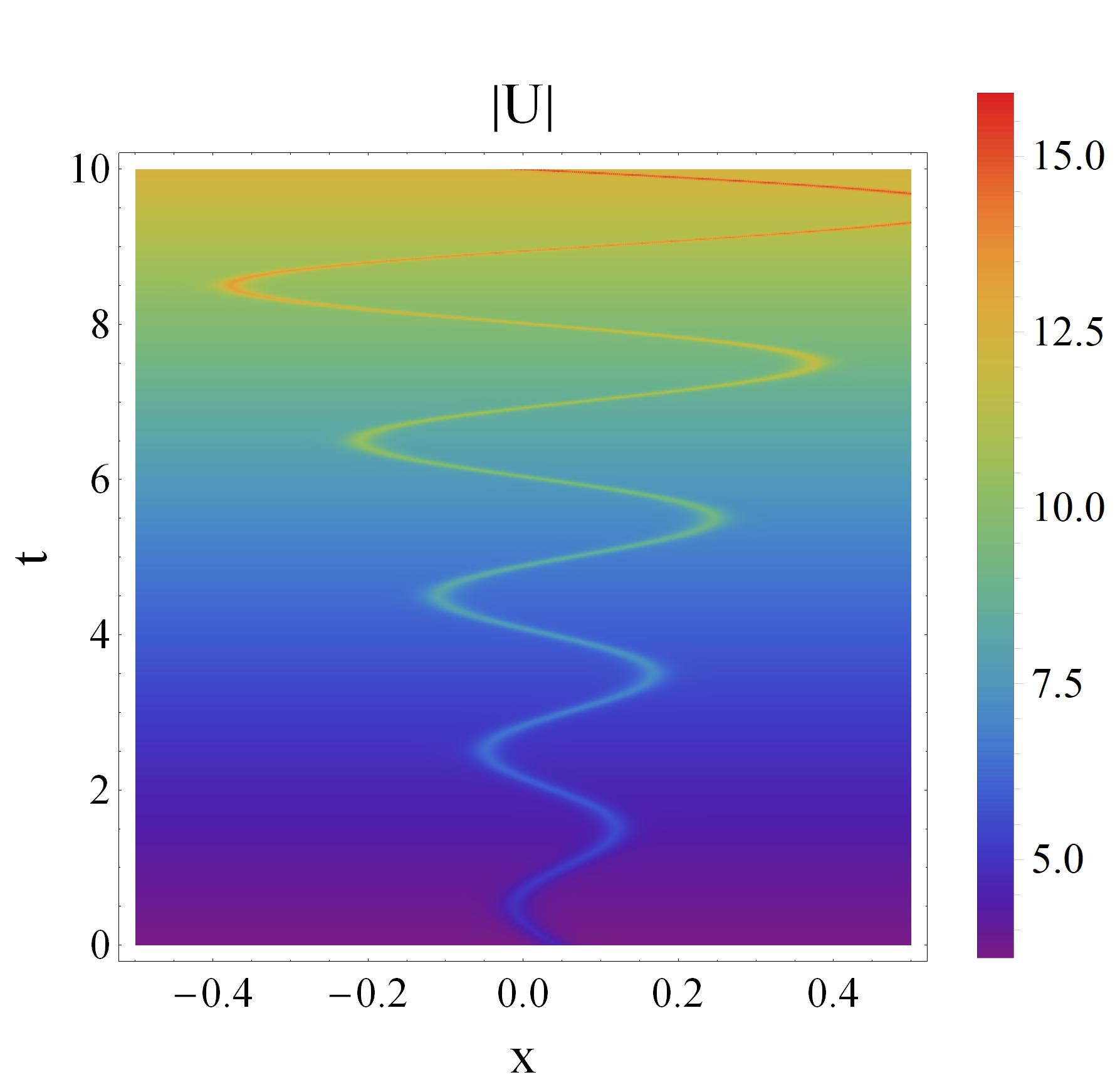}
		\caption{Anti-Dark 1SS}
		\label{Fig8e}
	\end{subfigure}
	\begin{subfigure}{.3\textwidth}
		\includegraphics[width=\textwidth]{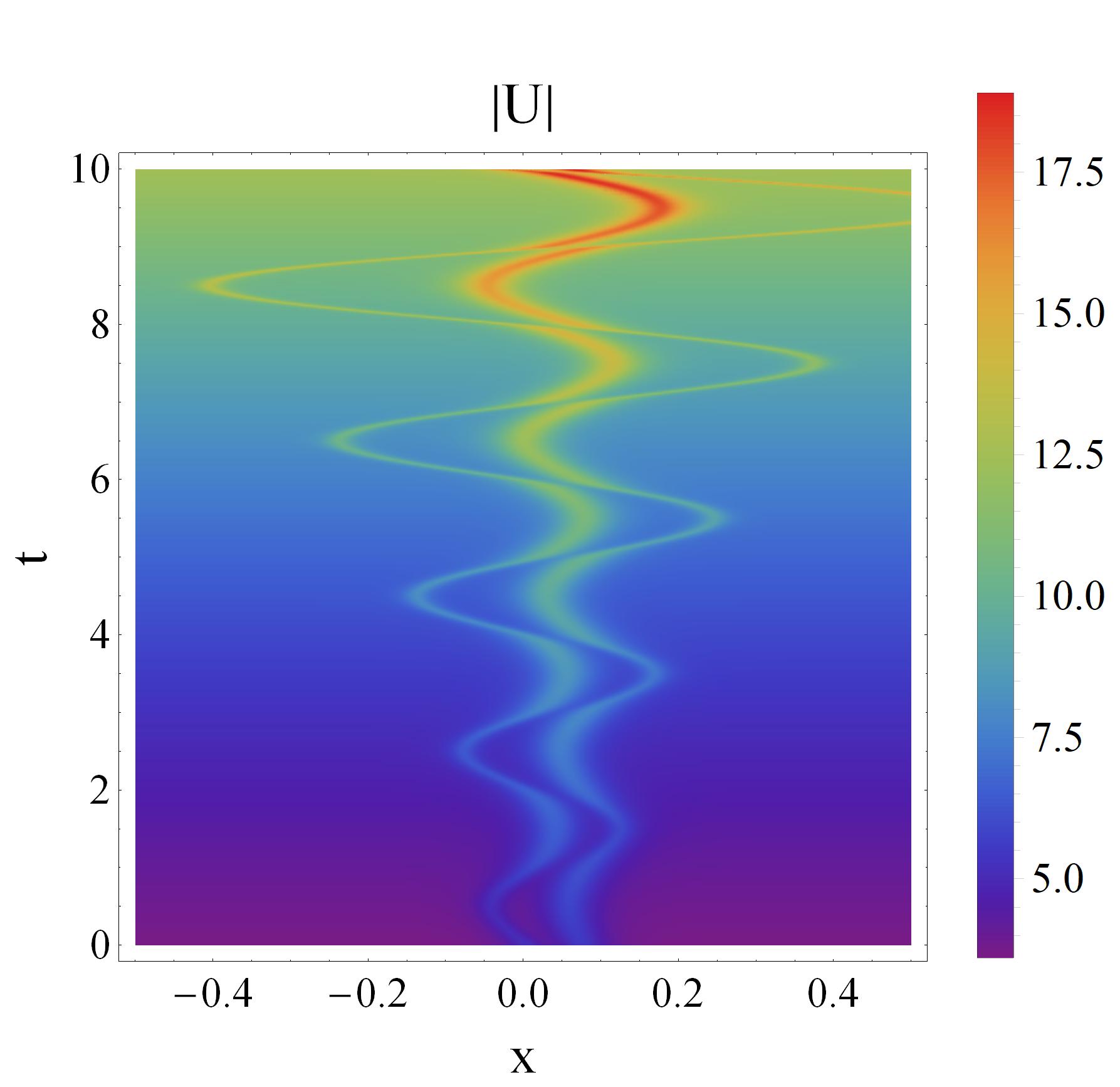}
		\caption{Anti-Dark 2SS}
		\label{Fig8f}
	\end{subfigure}
	\caption{Graphical representation of 2SS under SM scheme: $D(t) = e^{(\sigma t)}\ cos(k t)$ and $R(t) = cos(k t)$ and $\Gamma(t) = \frac{\sigma}{2}$ with $\sigma = 0.25$ and $k = \pi$. (a), (b) and (c) are 3D plot representation and (d), (e) and (f) are density plot representation. (a) and (d) represent anti-dark 1SS with $p_{1r} = -10$, $T_{1r} = -6$, $T_{1i}$ from Eqn. (\ref{cons}), (b) and (e) represent anti-dark 1SS with $p_{1r} = -20$, $T_{1r} = -9$, $T_{1i}$ from Eqn. (\ref{cons}) and (c) and (f) represent anti-dark-anti-dark 2SS with $p_{1r} = -10$, $p_{2r} = -20$, $T_{1r} = -6$, $T_{2r} = -9$, $T_{1i}$ and $T_{2i}$ from Eqn. (\ref{cons2}). In all the six graphs, we fix $\rho = 2$, $\kappa = 5$, $m = 0.2$, $n = 4$, $b = 1$, $a_2 = \frac{1}{n}$, $a_1 = m\ a_2$.}
	\label{Fig8}
\end{figure}

Now, for the SM scheme as demonstrated in subsection (\ref{subsubsec2.1.5}): $D(t) = e^{(\sigma t)}\ cos(k t)$, $R(t) = cos(k t)$ and hence $\Gamma(t) = \frac{\sigma}{2}$, we compare two 1SS represented in Figs. (\ref{Fig8a}) and (\ref{Fig8b}) with 2SS represented in Fig. (\ref{Fig8c}). We notice that upon interactions, the individual solitons of 2SS in Fig. (\ref{Fig8c}) repeatedly alternating their phases, thereby maintaining their original trajectories as they would have without interactions for 1SS case, as seen in Figs. (\ref{Fig8a}) and (\ref{Fig8b}). At a first glance of Fig. (\ref{Fig8c}), it might appear that the two individual solitons are not interacting with each other but in reality  they are interacting with repeated phase shifts that their trajectories appear to be of non interaction . Figs. (\ref{Fig8d}) - (\ref{Fig8f}) represent the same in density plot.

\subsection{Scheme for obtaining NSS}\label{subsec2.3}

In this subsection, we represent the scheme to obtain NSS of DLFLE and FLE. For this, we have to drop the terms greater than or equal to $\epsilon^{2N+1}$ in $g$, $f$ and $s$. Eqn. (\ref{bilin0}) becomes
\begin{align}
	\label{un}
	u &= \sqrt{\frac{D(\tau)}{R(\tau)}}\ \frac{g^{(N)}}{f^{(N)}}
\end{align}
the expansions of $g^{(N)}$ and $f^{(N)}$ are as follow
\begin{align}
	\label{gn}
	g^{(N)} &= g_0\ \Big(1 + \sum_{j=1}^N\ K_j\ e^{\theta_j + \theta_j^*}  + \frac{1}{2} \sum_{j,l=1;\ j\neq l}^N K_{jl}\ e^{\theta_j + \theta_j^* + \theta_l + \theta_l^*} + ... + K_{12...N}\ e^{\ \sum\limits_{j=1}^N \theta_j + \theta_j^*} \Big) \\
	\label{fn}
	f^{(N)} &= 1 + \sum_{j=1}^N\ T_j\ e^{\theta_j + \theta_j^*}  + \frac{1}{2} \sum_{j,l=1;\ j\neq l}^N T_{jl}\ e^{\theta_j + \theta_j^* + \theta_l + \theta_l^*} + ... + T_{12...N}\ e^{\ \sum\limits_{j=1}^N \theta_j + \theta_j^*}
\end{align}
where $\theta_j = p_j\ \xi + \Omega_j\ \tau$. $p_j$, $\Omega_j$, parameters within $g^{(N)}$ (i.e. $K's$) and within $f^{(N)}$ (i.e. $T's$) are complex parameters (subscript $j = 1$, ... , $N$). These parameters can be calculated using the bilinear Eqns. (\ref{BR1}) - (\ref{BR3}). Just like 2SS has two constraints (Eqn. (\ref{cons2})), the NSS consists of N number of constraints similar to that expressed in Eqn. (\ref{cons2}) but for $j = 1$, ... , $N$. The individual jth soliton of NSS has the same amplitude expression $A_j$ from Eqn. (\ref{A2}) and obey the same criteria for dark and anti-dark soliton nature with respect to the parameters $\kappa$, $p_{jr}$, $T_{ji}$ as already discussed for 1SS and 2SS. And eventually to obtain NSS of FLE, we have to put Eqn. (\ref{un}) into Eqn. (\ref{GT}).

\section{Gauge Connected Landau-Lifshitz equation}\label{sec3}

In this section we are going to establish the gauge connection between Eqn. (\ref{DLFLE}) with its counterpart LLE. To approach this, we exploit the equivalence of the Lax pairs of the two systems (DLFLE and LLE). Let us consider the Lax pair ($L$, $M$) for Eqn. (\ref{DLFLE})
\begin{align}
	\label{LaxFL}
	\Psi_{\xi} = L \Psi \quad \quad \quad \Psi_{\tau} = M \Psi
\end{align}
here $\Psi$ is the Jost function corresponding to the field function $u$ in which ($L$, $M$) are given by 
\begin{align}
	\label{LaxPairL}
	L &= \frac{-i\zeta^2 }{2} \Sigma  + \sqrt{ \frac{R(\tau)}{D(\tau)}}\ \zeta q_{\xi}          \\
	\label{LaxPairM}
	M &= \frac{i}{2 \zeta^2 } D(\tau) \Sigma - \frac{i}{\zeta} \sqrt{D(\tau)\ R(\tau)}\ \Sigma q + i R(\tau)\ \Sigma q^2
\end{align} 
$\zeta$ is the spectral parameter, $\Sigma$ and $q$ are $2\times 2 $ matrices given as follow

\vspace{4mm}
$ \quad \Sigma = \left(  \begin{tabular}{c c}
	1 & 0 \\ 
	0 & -1  \\ 
\end{tabular} \right), \quad \quad q = \left(  \begin{tabular}{c c}
	0 & u \\ 
	$-u^*$ & 0   \\
\end{tabular} \right) $ \\
from Eqn. (\ref{LaxFL}) follows the compatibility condition which is
\begin{align}
	\label{ComEqn}
	L_{\tau} - M_{\xi} + [L, M] = 0
\end{align}
Eqn. (\ref{ComEqn}) is also called zero curvature equation. Under a local gauge transformation we write a matrix $g$ such as $g({\xi}, {\tau}, \zeta_0) = \Psi ({\xi}, {\tau}, \zeta) |_{\zeta=\zeta_0}$ and the Jost function $\Psi$ transforms to $\Phi$ as $\quad \Psi \rightarrow \Phi ({\xi}, {\tau}, \zeta, \zeta_0) = g({\xi}, {\tau}, \zeta_0)^{-1} \Psi ({\xi}, {\tau}, \zeta)$. $\Phi$ is the Jost function corresponding to the spin field function $S$ of the LL system. The new Lax pair ($L^\prime$, $M^\prime$) associated with $\Phi$ written as
\begin{align}
	\label{LaxFL2}
	\Phi_{\xi} = L^\prime \Phi \quad \quad \quad \Phi_{\tau} = M^\prime \Phi	
\end{align}
and the relation between ($L^\prime$, $M^\prime$) and ($L$, $M$) expressed as
\begin{align}
	\label{LaxPairL2}
	L^\prime = g^{-1} L g - g^{-1} g_{\xi} \\
	\label{LaxPairM2}
	M^\prime = g^{-1} M g - g^{-1} g_{\tau}
\end{align}
where
\begin{align}
	\label{gxgt}
	g_{\xi} g^{-1} = L |_{\zeta = \zeta_0}, \quad \quad \quad  g_{\tau} g^{-1} = M |_{\zeta = \zeta_0}
\end{align}
the compatibility condition Eqn. (\ref{ComEqn}) for the new system rewritten as
\begin{align}
	\label{ComEqn2}
	L^\prime_{\tau} - M^\prime_{\xi} + [L^\prime, M^\prime] = 0
\end{align}
Eqn. (\ref{ComEqn2}) is corresponding to the spin field function ($S$). $S$ is expressed as
\begin{align}
	\label{S}
	S = g^{-1} \Sigma g
\end{align}
and $S^2 = I$. This leads to the expressions for the derivatives of $S$ namely $S_{\xi}$, $S_{\tau}$ and $S_{\tau \xi}$ as
\begin{align}
	S_{\xi} = g^{-1} [\Sigma, L_0] g
	\label{Sx}
	\Rightarrow S_{\xi} = 2 \zeta_0 \sqrt{\frac{R(\tau)}{D(\tau)}}\ g^{-1} \Sigma q_{\xi} g\\
	S_{\tau} = g^{-1} [\Sigma, M_0] g
	\label{St}
	\Rightarrow S_{\tau} = - \frac{2 i}{\zeta_0} \sqrt{D(\tau) R(\tau)}\ g^{-1} q g\\
	\label{Stx}
	S_{\tau \xi} = 2i R(\tau) (g^{-1} q_{\xi}\ q g - g^{-1} q\ q_{\xi} g) - \frac{2i}{\zeta_0} \sqrt{D(\tau) R(\tau)}\ g^{-1} q_{\xi} g
\end{align}
using the above equations, the compatibility condition Eqn. (\ref{ComEqn2}) yields
\begin{align}
	\label{Gauge}
	D(\tau)\ S_{\xi} + \frac{\zeta_0^2}{2 i}\ [S, S_{{\tau}{\xi}}] = 0 
\end{align}
This is the gauge connected LLE for the Eqn. (\ref{DLFLE}). One important thing to note is that Eqn. (\ref{Gauge}) is consistent for the system of Eqns. (\ref{LaxFL2}), (\ref{LaxPairL2}), (\ref{LaxPairM2}) in any matrix $S$ of arbitrary dimension as long as $S$ satisfies $S^2 = I$.


\section{Conclusion}

We have transformed FLE under SM into DLFLE and using Hirota bilinearization we obtained 1SS and 2SS of DLFLE and under reverse transformation we obtained 1SS and 2SS of FLE. Under different SM scheme, we have explored the soliton using graphical representation. We have also mentioned that under certain SM scheme the soliton can be amplified accompanied by a certain amount of affecting the background. This information can be utilized to construct a desirable FLE system for ultrashort pulse with customized properties. Thereafter, we have determined the shift in phase of the individual solitons suffered upon interaction and also showed that the interaction was elastic. We believe that the derived soliton solutions will be a useful information carrier in digital and quantum communication where higher order effects like STD and ND are taken into account along with GVD, KLNE. The proposed method may be extended to other nonlinear models such as coupled FLE under SM to obtain soliton solutions in multi-mode waveguide. This will be one of our future objectives. Secondly, we have obtained a gauge equivalence between DLFLE and LL spin system with the explicit construction of the new equivalent Lax pair. LLE exhibits a variety of nonlinear structures like spin waves, solitary wave, soliton, spatio-temporal patterns, etc. The derived LLE can be useful to study analogous properties of FLE in a spin system.

\bmhead{Acknowledgments}

R. Dutta and S. Talukdar acknowledge Department of Science \& Technology, Govt. of India for INSPIRE fellowship award. Corresponding award numbers DST/INSPIRE Fellowship/2020/IF200303 and DST/INSPIRE Fellowship/2020/IF200278.

\bibliography{FLESM-bibliography}

\begin{thebibliography}{45}
\providecommand{\natexlab}[1]{#1}
\providecommand{\url}[1]{{#1}}
\providecommand{\urlprefix}{URL }
\providecommand{\doi}[1]{\url{https://doi.org/#1}}
\providecommand{\eprint}[2][]{\url{#2}}
 \bibcommenthead

\bibitem[{Baronio et~al(2015)Baronio, Chen, Grelu, Wabnitz, and
  Conforti}]{baronio2015baseband}
Baronio F, Chen S, Grelu P, et~al (2015) Baseband modulation instability as the
  origin of rogue waves. Physical Review A 91(3):033804

\bibitem[{Berger and Perkins(1976)}]{berger1976thresholds}
Berger R, Perkins F (1976) Thresholds of parametric instabilities near the
  lower-hybrid frequency. The Physics of Fluids 19(3):406--411

\bibitem[{Biswas et~al(2018)Biswas, Y{\i}ld{\i}r{\i}m, Ya{\c{s}}ar, Zhou,
  Moshokoa, and Belic}]{biswas2018optical}
Biswas A, Y{\i}ld{\i}r{\i}m Y, Ya{\c{s}}ar E, et~al (2018) Optical soliton
  solutions to fokas-lenells equation using some different methods. Optik
  173:21--31

\bibitem[{Chakraborty et~al(2015)Chakraborty, Nandy, and
  Barthakur}]{chakraborty2015bilinearization}
Chakraborty S, Nandy S, Barthakur A (2015) Bilinearization of the generalized
  coupled nonlinear schr{\"o}dinger equation with variable coefficients and
  gain and dark-bright pair soliton solutions. Physical Review E 91(2):023210

\bibitem[{Chen et~al(2018)Chen, Ye, Soto-Crespo, Grelu, and
  Baronio}]{chen2018peregrine}
Chen S, Ye Y, Soto-Crespo JM, et~al (2018) Peregrine solitons beyond the
  threefold limit and their two-soliton interactions. Physical Review Letters
  121(10):104101

\bibitem[{Cinar et~al(2022)Cinar, Secer, Ozisik, and
  Bayram}]{cinar2022derivation}
Cinar M, Secer A, Ozisik M, et~al (2022) Derivation of optical solitons of
  dimensionless fokas-lenells equation with perturbation term using sardar
  sub-equation method. Optical and Quantum Electronics 54(7):402

\bibitem[{Davydova and Lashkin(1991)}]{davydova1991short}
Davydova T, Lashkin V (1991) Short-wavelength ion-cyclotron soliton. Soviet
  Journal of Plasma Physics 17(8):568--570

\bibitem[{Dutta et~al(2023)Dutta, Talukdar, Saharia, and
  Nandy}]{dutta2023fokas}
Dutta R, Talukdar S, Saharia GK, et~al (2023) Fokas-lenells equation dark
  soliton and gauge equivalent spin equation. Optical and Quantum Electronics
  55(13):1183

\bibitem[{El-Shiekh and Hamdy(2023)}]{el2023novel}
El-Shiekh RM, Hamdy H (2023) Novel distinct types of optical solitons for the
  coupled fokas-lenells equations. Optical and Quantum Electronics 55(3):251

\bibitem[{Gaballah et~al(2023)Gaballah, El-Shiekh, and
  Hamdy}]{gaballah2023generalized}
Gaballah M, El-Shiekh RM, Hamdy H (2023) Generalized periodic and soliton
  optical ultrashort pulses for perturbed fokas--lenells equation. Optical and
  Quantum Electronics 55(4):364

\bibitem[{Ghosh and Nandy(1999)}]{ghosh1999inverse}
Ghosh S, Nandy S (1999) Inverse scattering method and vector higher order
  non-linear schr{\"o}dinger equation. Nuclear Physics B 561(3):451--466

\bibitem[{Ghosh et~al(1999)Ghosh, Kundu, and Nandy}]{ghosh1999soliton}
Ghosh S, Kundu A, Nandy S (1999) Soliton solutions, liouville integrability and
  gauge equivalence of sasa satsuma equation. Journal of Mathematical Physics
  40(4):1993--2000

\bibitem[{Gomez~S et~al(2022)Gomez~S, Roshid, Inc, Akinyemi, and
  Rezazadeh}]{gomez2022soliton}
Gomez~S CA, Roshid HO, Inc M, et~al (2022) On soliton solutions for perturbed
  fokas--lenells equation. Optical and Quantum Electronics 54(6):370

\bibitem[{Gonz{\'a}lez-Gaxiola and Biswas(2018)}]{gonzalez2018w}
Gonz{\'a}lez-Gaxiola O, Biswas A (2018) W-shaped optical solitons of
  chen--lee--liu equation by laplace--adomian decomposition method. Optical and
  Quantum Electronics 50:1--11

\bibitem[{Guo and Ding(2008)}]{guo2008landau}
Guo B, Ding S (2008) Landau-Lifshitz Equations, vol~1. World Scientific

\bibitem[{Hasegawa and Tappert(1973{\natexlab{a}})}]{hasegawa1973transmission}
Hasegawa A, Tappert F (1973{\natexlab{a}}) Transmission of stationary nonlinear
  optical pulses in dispersive dielectric fibers. i. anomalous dispersion.
  Applied Physics Letters 23(3):142--144

\bibitem[{Hasegawa and Tappert(1973{\natexlab{b}})}]{hasegawa1973transmission2}
Hasegawa A, Tappert F (1973{\natexlab{b}}) Transmission of stationary nonlinear
  optical pulses in dispersive dielectric fibers. ii. normal dispersion.
  Applied Physics Letters 23(4):171--172

\bibitem[{Hirota and Satsuma(1976)}]{hirota1976n}
Hirota R, Satsuma J (1976) N-soliton solutions of model equations for shallow
  water waves. Journal of the Physical Society of Japan 40(2):611--612

\bibitem[{Hosseini et~al(2020)Hosseini, Mirzazadeh, Vahidi, and
  Asghari}]{hosseini2020optical}
Hosseini K, Mirzazadeh M, Vahidi J, et~al (2020) Optical wave structures to the
  fokas--lenells equation. Optik 207:164450

\bibitem[{Hosseini et~al(2022)Hosseini, Sadri, Salahshour, Baleanu, Mirzazadeh,
  and Inc}]{hosseini2022generalized}
Hosseini K, Sadri K, Salahshour S, et~al (2022) The generalized sasa--satsuma
  equation and its optical solitons. Optical and Quantum Electronics 54(11):723

\bibitem[{Hosseini et~al(2023)Hosseini, Hincal, Salahshour, Mirzazadeh,
  Dehingia, and Nath}]{hosseini2023dynamics}
Hosseini K, Hincal E, Salahshour S, et~al (2023) On the dynamics of soliton
  waves in a generalized nonlinear schr{\"o}dinger equation. Optik 272:170215

\bibitem[{Jawad et~al(2019)Jawad, Al~Azzawi, Biswas, Khan, Zhou, Moshokoa, and
  Belic}]{jawad2019bright}
Jawad AJM, Al~Azzawi FJI, Biswas A, et~al (2019) Bright and singular optical
  solitons for kaup--newell equation with two fundamental integration norms.
  Optik 182:594--597

\bibitem[{Krishnan et~al(2019)Krishnan, Biswas, Zhou, and
  Alfiras}]{krishnan2019optical}
Krishnan E, Biswas A, Zhou Q, et~al (2019) Optical soliton perturbation with
  fokas--lenells equation by mapping methods. Optik 178:104--110

\bibitem[{Kundu(1984)}]{kundu1984landau}
Kundu A (1984) Landau--lifshitz and higher-order nonlinear systems gauge
  generated from nonlinear schr{\"o}dinger-type equations. Journal of
  mathematical physics 25(12):3433--3438

\bibitem[{Kundu(2010)}]{kundu2010two}
Kundu A (2010) Two-fold integrable hierarchy of nonholonomic deformation of the
  derivative nonlinear schr{\"o}dinger and the lenells--fokas equation. Journal
  of mathematical physics 51(2)

\bibitem[{Lakshmanan(2011)}]{lakshmanan2011fascinating}
Lakshmanan M (2011) The fascinating world of the landau--lifshitz--gilbert
  equation: an overview. Philosophical Transactions of the Royal Society A:
  Mathematical, Physical and Engineering Sciences 369(1939):1280--1300

\bibitem[{Lashkin(2021)}]{lashkin2021perturbation}
Lashkin V (2021) Perturbation theory for solitons of the fokas-lenells
  equation: Inverse scattering transform approach. Physical Review E
  103(4):042203

\bibitem[{Lenells(2009)}]{lenells2009exactly}
Lenells J (2009) Exactly solvable model for nonlinear pulse propagation in
  optical fibers. Studies in Applied Mathematics 123(2):215--232

\bibitem[{Lenells and Fokas(2008)}]{lenells2008novel}
Lenells J, Fokas A (2008) On a novel integrable generalization of the nonlinear
  schr{\"o}dinger equation. Nonlinearity 22(1):11

\bibitem[{Liu et~al(2017)Liu, Tian, Chai, and Yuan}]{liu2017certain}
Liu L, Tian B, Chai HP, et~al (2017) Certain bright soliton interactions of the
  sasa-satsuma equation in a monomode optical fiber. Physical Review E
  95(3):032202

\bibitem[{L{\"u} and Peng(2013)}]{lu2013nonautonomous}
L{\"u} X, Peng M (2013) Nonautonomous motion study on accelerated and
  decelerated solitons for the variable-coefficient lenells-fokas model. Chaos:
  An Interdisciplinary Journal of Nonlinear Science 23(1)

\bibitem[{Malomed and Weinstein(1996)}]{malomed1996soliton}
Malomed B, Weinstein MI (1996) Soliton dynamics in the discrete nonlinear
  schr{\"o}dinger equation. Physics Letters A 220(1-3):91--96

\bibitem[{Matsuno(2012{\natexlab{a}})}]{matsuno2012direct1}
Matsuno Y (2012{\natexlab{a}}) A direct method of solution for the
  fokas--lenells derivative nonlinear schr{\"o}dinger equation: I. bright
  soliton solutions. Journal of Physics A: Mathematical and Theoretical
  45(23):235202

\bibitem[{Matsuno(2012{\natexlab{b}})}]{matsuno2012direct}
Matsuno Y (2012{\natexlab{b}}) A direct method of solution for the
  fokas--lenells derivative nonlinear schr{\"o}dinger equation: Ii. dark
  soliton solutions. Journal of Physics A: Mathematical and Theoretical
  45(47):475202

\bibitem[{Onder et~al(2022)Onder, Secer, Ozisik, and
  Bayram}]{onder2022obtaining}
Onder I, Secer A, Ozisik M, et~al (2022) Obtaining optical soliton solutions of
  the cubic--quartic fokas--lenells equation via three different analytical
  methods. Optical and Quantum Electronics 54(12):786

\bibitem[{Pethick and Smith(2008)}]{pethick2008bose}
Pethick CJ, Smith H (2008) Bose--Einstein condensation in dilute gases.
  Cambridge university press

\bibitem[{Serkin and Hasegawa(2000)}]{serkin2000novel}
Serkin VN, Hasegawa A (2000) Novel soliton solutions of the nonlinear
  schr{\"o}dinger equation model. Physical Review Letters 85(21):4502

\bibitem[{Takhtajan and Zakharov(1979)}]{takhtajan1979equivalence}
Takhtajan L, Zakharov V (1979) Equivalence of the nonlinear schrodinger
  equation and the heisenbergferromagnet equation. Theor Math Phys 38:17--23

\bibitem[{Talukdar et~al(2023)Talukdar, Dutta, Saharia, and
  Nandy}]{talukdar2023multi}
Talukdar S, Dutta R, Saharia GK, et~al (2023) Multi soliton solutions of the
  fokas--lenells equation using modified bilinear method and conservation laws.
  Journal of Optics pp 1--9

\bibitem[{Triki and Biswas(2018)}]{triki2018sub}
Triki H, Biswas A (2018) Sub pico-second chirped envelope solitons and
  conservation laws in monomode optical fibers for a new derivative nonlinear
  schr{\"o}dinger's model. Optik 173:235--241

\bibitem[{Triki and Wazwaz(2017)}]{triki2017combined}
Triki H, Wazwaz AM (2017) Combined optical solitary waves of the
  fokas—lenells equation. Waves in Random and Complex Media 27(4):587--593

\bibitem[{Ullah et~al(2023)Ullah, Seadawy, and Ali}]{ullah2023optical}
Ullah MS, Seadawy AR, Ali MZ (2023) Optical soliton solutions to the
  fokas--lenells model applying the $\varphi$ 6-model expansion approach.
  Optical and Quantum Electronics 55(6):495

\bibitem[{Wang et~al(2017)Wang, Wang, Wang, Sun, and Qi}]{wang2017higher}
Wang ZQ, Wang X, Wang L, et~al (2017) Higher-order peregrine combs and
  peregrine walls for the variable-coefficient lenells-fokas equation.
  Superlattices and Microstructures 102:189--201

\bibitem[{Yildirim(2019)}]{yildirim2019opticalss}
Yildirim Y (2019) Optical solitons to sasa--satsuma model with trial equation
  approach. Optik 184:70--74

\bibitem[{Y{\i}ld{\i}r{\i}m et~al(2020)Y{\i}ld{\i}r{\i}m, Biswas, Asma, Ekici,
  Ntsime, Zayed, Moshokoa, Alzahrani, and Belic}]{yildirim2020optical}
Y{\i}ld{\i}r{\i}m Y, Biswas A, Asma M, et~al (2020) Optical soliton
  perturbation with chen--lee--liu equation. Optik 220:165177

\end{thebibliography}

\section*{Statements \& Declarations}

\bmhead{Funding}

R. Dutta and S. Talukdar receive fellowship from Department of Science \& Technology, Govt. of India under INSPIRE programme. Corresponding fellowship award numbers DST/INSPIRE Fellowship/2020/IF200303 and DST/INSPIRE Fellowship/2020/IF200278. Other than this, the authors did not receive any fund support from any organization for the submitted work.

\bmhead{Competing Interests}

The authors have no relevant financial or non-financial interests to disclose.

\bmhead{Authors Contribution}

Conceptualization: Riki Dutta, Sudipta Nandy; Formal Analysis: Riki Dutta; Supervision: Sudipta Nandy; Validation: Sudipta Nandy, Sagardeep Talukdar, Gautam K. Saharia; Writing - original draft: Riki Dutta; Writing - review \& editing: All the authors.

\end{document}